\documentclass[preprint,hidelinks]{elsarticle}

\makeatletter
\def\ps@pprintTitle{%
 \let\@oddhead\@empty
 \let\@evenhead\@empty
 \def\@oddfoot{}%
 \let\@evenfoot\@oddfoot}
\makeatother
\usepackage{latexsym}
\usepackage{amssymb}
\usepackage{amsfonts}
\usepackage{amstext}
\usepackage{amsmath}
\usepackage{multicol}
\usepackage{xcolor}
\usepackage{array}
\usepackage{hyperref}
\usepackage{mathtools}
\usepackage{geometry}
\usepackage{bm}
\usepackage{tikz-cd}
\usepackage{graphicx}
\usepackage{epstopdf}
\usepackage[hypcap=false]{caption}
\usepackage{subcaption}
\usepackage{pstricks}
\usepackage{booktabs}
\usepackage{natbib}
\usepackage{starfont}
\hypersetup{colorlinks=true,linkcolor=black,urlcolor=blue,citecolor=blue}

\graphicspath{{figs/}}

\newcommand{\bb}{\mathbf{b}}

\usepackage{calc}
\newlength\lone

\definecolor{greenx}{rgb}{0.4660,0.6740,0.1880}
\definecolor{purpdot}{rgb}{0.4940,0.1840,0.5560}

\newcommand\solidrule[1][15pt]{\rule[0.5ex]{#1}{1pt}}
\newcommand\dashedrule{\mbox{%
	\solidrule[3pt]\hspace{3pt}\solidrule[3pt]\hspace{3pt}\solidrule[3pt]}}

\newcommand\dottedrule{\mbox{%
	\solidrule[1.5pt]\hspace{2pt}\solidrule[1.5pt]\hspace{2pt}\solidrule[1.5pt]\hspace{2pt}\solidrule[1.5pt]\hspace{2pt}\solidrule[1.5pt]}}

\newcommand{\R}{\mathcal{R}}
\newcommand{\ov}{\overline}
\newcommand{\bx}{\mathbf{x}}
\newcommand{\Q}{\mathcal{Q}}
\newcommand{\eps}{\varepsilon}

\begin{document}

\begin{frontmatter}

\title{Heat transport bounds for a truncated model of Rayleigh--B\'enard convection via polynomial optimization}

\author[add1]{Matthew L. Olson\corref{cor1}}
\ead{mlolson@umich.edu}
\author[add2]{David Goluskin}
\ead{goluskin@uvic.ca}
\author[add3,add4]{William W. Schultz}
\ead{schultz@umich.edu}
\author[add1,add5,add6]{Charles R. Doering}
\ead{doering@umich.edu}

\cortext[cor1]{Corresponding Author.}
\address[add1]{Department of Mathematics, University of Michigan, Ann Arbor, MI 48109, USA}
\address[add2]{Department of Mathematics \& Statistics, University of Victoria, Victoria, BC V8P 5C2, Canada}
\address[add3]{Department of Mechanical Engineering, University of Michigan, Ann Arbor, MI 48109, USA}
\address[add4]{Department of Naval Architecture \& Marine Engineering, University of Michigan, Ann Arbor, MI 48109, USA}
\address[add5]{Center for the Study of Complex Systems, University of Michigan, Ann Arbor, MI 48109, USA}
\address[add6]{Department of Physics, University of Michigan, Ann Arbor, MI 48109, USA}

\begin{abstract}
Upper bounds on time-averaged heat transport are obtained for an eight-mode Galerkin truncation of Rayleigh's 1916 model of natural thermal convection. Bounds for the ODE model---an extension of Lorenz's three-ODE system---are derived by constructing auxiliary functions that satisfy sufficient conditions wherein certain polynomial expressions must be nonnegative. Such conditions are enforced by requiring the polynomial expressions to admit sum-of-squares representations, allowing the resulting bounds to be minimized using semidefinite programming. Sharp or nearly sharp bounds on mean heat transport are computed numerically for numerous values of the model parameters: the Rayleigh and Prandtl numbers and the domain aspect ratio. In all cases where the Rayleigh number is small enough for the ODE model to be quantitatively close to the PDE model, mean heat transport is maximized by steady states. In some cases at larger Rayleigh number, time-periodic states maximize heat transport in the truncated model. Analytical parameter-dependent bounds are derived using quadratic auxiliary functions, and they are sharp for sufficiently small Rayleigh numbers.
\end{abstract}


\end{frontmatter}

\section{Introduction} \label{sec:intro}
Rayleigh--B\'enard convection models the fundamental physics of buoyancy-driven flow in a fluid layer heated from below.
Rayleigh's seminal model~\cite{Rayleigh1916}, comprised of the Boussinesq approximation to the Navier--Stokes equations in an idealized two-dimensional domain with stress-free isothermal boundaries, has been a primary paradigm of nonlinear dynamics for over a century.
Its study has inspired significant theoretical advances on many fronts, including linear~\cite{Rayleigh1916} and nonlinear~\cite{Joseph1976} stability analysis, weakly nonlinear theory~\cite{MalkusVeronis1958}, pattern formation~\cite{NewellWhitehead1969}, and chaos~\cite{Lorenz63}.

The relationship between the magnitude of the imposed temperature gradient and the resulting rate of heat transport---characterizing the effective thermal conductivity of the convecting layer---is of particular importance for many applications in engineering and the applied sciences.
Transport is quantified by the Nusselt number, $\rm{Nu}$, the average convective enhancement of heat flux over purely conductive flux. Typically one seeks the dependence of $\rm{Nu}$ on one or more model parameters, such as the Rayleigh number, $\rm{Ra}$, a dimensionless parameter that is proportional to the imposed temperature gradient. Flows maximizing ${\rm Nu}$ might be turbulent but need not be; they could be steady or time-periodic, and dynamically stable or unstable.
One motivation for this study is to investigate whether simple steady states maximize heat transport, an idea suggested by recent computations of steady solutions with isothermal no-slip boundaries that transport heat at slightly higher rates than three-dimensional turbulent flows~\cite{Sondak2015, Waleffe2015,Wen2020noslip}.

Deducing the $\rm{Nu}$--$\rm{Ra}$ relation is challenging in part because coexisting solutions in the strongly nonlinear regime may transfer heat at widely varying rates~\cite{Goluskin2013, GJFS2014JFM}.
Hence it is natural to focus analysis on bounding convective heat transport among all possible solutions.
Determining upper bounds on transport in Rayleigh's model and variations thereof has been an active area of inquiry for well over a half century~\cite{Busse69, Doering96, Howard63, WCKD2015PRE, WD2011PRL}.
The best known upper bounds on heat transport in Rayleigh's two-dimensional configuration with isothermal stress-free boundaries take the form ${\rm{Nu}\le0.106\,Ra^{5/12}}$ in the limit of large $\rm{Ra}$~\cite{WCKD2015PRE, Whitehead2011}.
It is not known whether there exist solutions achieving the growth rate of this bound; no solutions have been found where Nu grows faster than ${\cal O}(\rm{Ra}^{1/3})$ as $\rm{Ra} \rightarrow \infty$ \cite{Chini2009,Wen2020}.

A bound on ${\rm Nu}$ at particular parameter values is sharp if and only if it is attained by a solution of the equations of motion. Confirming that a particular steady flow maximizes $\rm{Nu}$ requires proving an upper bound with the same value. When this occurs we say that the maximizing solution saturates the upper bound. Perfectly sharp bounds are typically beyond reach for nonlinear partial differential equations (PDEs) like the Boussinesq equations governing Rayleigh--B\'enard convection. For ordinary differential equations (ODEs), however, the recent development of methods based on polynomial optimization has made the construction of sharp bounds tractable. In this paper we employ such methods to construct upper bounds on the finite-dimensional analogue of the Nusselt number for a distinguished eight-ODE truncation of the Boussinesq equations. We compute \emph{auxiliary functions} that by satisfying certain inequalities imply bounds on time averages, similar to how Lyapunov functions imply stability of a particular state in a dynamical system.

Given any well-posed ODE with bounded trajectories $\bx(t)$, there exist continuously differentiable auxiliary functions producing arbitrarily sharp bounds on the time average of any continuous $\Phi(\mathbf x)$~\cite{Tobasco2018}.
The practical challenge is to construct the auxiliary functions that satisfy the suitable inequalities.
This is analytically intractable in general, but it can be implemented with computer assistance when the phase space vector field $\mathbf f(\mathbf x)$ and quantity of interest $\Phi(\mathbf x)$ are polynomial and one seeks polynomial auxiliary functions. 
In such cases the relevant inequality conditions correspond to nonnegativity of certain polynomial expressions; this nonnegativity can be enforced by requiring the polynomials to admit sum-of-squares (SOS) representations.
Optimization over polynomials subject to SOS constraints can be carried out computationally as a semidefinite program (SDP).

The use of SOS optimization to bound time averages was proposed in~\cite{Cherny2014} and has since been applied to various dynamical systems~\cite{Fantuzzi2016, Goluskin2018, Goluskin2019}. In the  present work we apply this technique to compute upper bounds on time-averaged heat transport for a truncated model of Rayleigh--B\'enard convection. Upper bounds are constructed across a wide range of the three dimensionless model parameters: the Rayleigh and Prandtl numbers and the domain aspect ratio. The Nusselt number is also computed along various particular solutions of the ODEs to identify solutions whose heat transport saturates the bounds.

The rest of this paper is organized as follows.
The governing PDEs and the truncated ODE model that we analyze are presented in~\S\ref{sec:RBC} and~\S\ref{sec:ROM}, respectively.
Various particular solutions of this system of ODEs are presented in~\S\ref{sec:particular}, providing candidates for the maximal heat transport that can be compared to the upper bounds.
Section~\ref{sec:SOS} describes the general construction of bounds on time-averaged quantities for ODEs using SOS optimization. In~\S\ref{sec:Numerical SDP}, computational constructions for our particular ODE model are presented using polynomial auxiliary functions up to degree eight, giving very tight numerical bounds on the heat transport.
The auxiliary function approach is carried out analytically in~\S\ref{sec:Quad} for quadratic auxiliary functions, yielding bounds that are weaker than the numerical bounds but have explicit parameter dependence.
Conclusions are presented in~\S\ref{sec:Con}, followed by computational and analytical details in the appendices.

\section{Rayleigh--B\'enard convection} \label{sec:RBC}
In Rayleigh's 1916 model~\cite{Rayleigh1916}, a two-dimensional incompressible fluid lies between stress-free top and bottom boundaries. A temperature gradient in the fluid is maintained by heating the bottom boundary to a higher temperature than the top one. The velocity $\mathbf{u}(x,z,t)$, pressure $p(x,z,t)$, and temperature $T(x,z,t)$ are governed by the Boussinesq approximation to the Navier--Stokes equations, expressed in dimensionless form as 
\begin{equation} \label{eq:NSE}
\begin{aligned}
\partial_t \mathbf{u} + \mathbf{u} \cdot \nabla \mathbf{u} &= - \nabla p + \sigma \nabla^2 \mathbf{u} + \sigma \text{Ra} \,T \hat{\mathbf{z}}, \\ 
\nabla \cdot \mathbf{u} &= 0, \\
\partial_t T + \mathbf{u} \cdot \nabla T &= \nabla^2 T.
\end{aligned}
\end{equation}
To construct the dimensionless equations \eqref{eq:NSE}, the length scale $d$ is chosen so that the dimensional layer height is $\pi d$, time is nondimensionalized using the thermal diffusion time $d^2/\kappa$, and the temperature scale is given by the dimensional temperature drop $\Delta$ from the bottom to the top boundary. We define the temperature such that $T = 1$ along the bottom, implying that $T=0$ along the top one. The remaining material parameters---the thermal diffusivity $\kappa$, kinematic viscosity $\nu$, gravitational acceleration $g$ in the $-z$ direction, and linear coefficient of thermal expansion $\alpha$---form two dimensionless groups, typically chosen as the Prandtl number $\sigma = \frac{\nu}{\kappa}$ and Rayleigh number $\rm{Ra} = \frac{g \alpha \Delta (\pi d)^3}{\kappa \nu}$.  It is convenient to define a modified Rayleigh number $\R := \mbox{Ra}/\pi^4$ to avoid extra factors of $\pi$ in the dimensionless form.

We consider convection in the dimensionless rectangular domain $(x,z) \in [0, \pi A] \times [0,\pi]$ of aspect ratio $A$ that is periodic in the horizontal coordinate $x$. The equations of motion in two spatial dimensions can be written in terms of the stream function $\psi(x,z,t)$, defined such that the horizontal and vertical velocity components are given by $(u,w) = (\partial_z \psi, - \partial_x \psi)$.
Finally, the dimensionless negative temperature deviation is defined as $\theta := \pi \R(T_c - T)$, where $T_c := 1 - z/\pi$ is the dimensionless linear temperature profile of the purely conducting state.

 In terms of $\psi$ and $\theta$ the dimensionless Boussinesq equations are
\begin{equation} 
\label{eq:BE}
\begin{aligned}
\partial_t \nabla^2 \psi - \{\psi,\nabla^2 \psi \}& = \sigma \nabla^4 \psi + \sigma \partial_x \theta, \\
\partial_t \theta - \{\psi,\theta \} &= \nabla^2 \theta + \mathcal{R} \partial_x \psi,
\end{aligned}
\end{equation}
where $\{f,g\} := \partial_x f \, \partial_z g - \partial_z f \, \partial_x g$ denotes the Poisson bracket.

Stress-free isothermal boundary conditions require $\partial_z^2\psi$ and $\theta$ to vanish at the top and bottom boundaries. Impenetrability requires $\psi$ to be constant on both boundaries, and in a zero-momentum reference frame both constants can be chosen to be zero without loss of generality. Thus, the boundary conditions on $\psi$ and $\theta$ are
\begin{equation}\label{eq:BCs}
\psi,~\partial_{z}^2\psi,~\theta = 0 \quad \text{at}\quad z=0,\pi.
\end{equation}
No-slip boundary conditions, corresponding to $\partial_{z}\psi$ rather than $\partial_{z}^2\psi$ vanishing on the impenetrable boundaries, are of interest as well but are not suitable for Fourier expansion. Following Rayleigh we may content ourselves with stress-free boundaries where, as explained in the next section, expansions of $\psi$ and $\theta$ in familiar Fourier basis functions readily produce models with desirable properties.

The Nusselt number quantifying convective transport is defined as the ratio of total vertical heat flux, averaged over volume and infinite time, to conductive flux. To define ${\rm Nu}$ in terms of $\psi$ and $\theta$, let the spatial average over the domain $(x,z) \in [0,A\pi]\times [0,\pi]$ be denoted as
\begin{equation} \label{eq:spaceAvg}
\langle f \rangle := \frac{1}{A\pi^2} \int_0^\pi \int_0^{A\pi} f(x,z) \, {\rm d}x \, {\rm d}z,
\end{equation}
and denote the infinite-time average as
\begin{equation}  \label{eq:timeAvg}
\ov{f} := \lim_{\tau \to \infty} \frac{1}{\tau} \int_0^\tau f(t) \, {\rm d}t.
\end{equation}
To ensure limits exist one may instead define time averages using a limsup or liminf. The Nusselt number along a particular solution to \eqref{eq:BE} is given by \cite{Goluskin2015a}
\begin{equation} \label{eq:NuDef}
{\rm Nu} = 1 + \tfrac{1}{\mathcal{R}} \ov{\langle \theta \partial_x \psi \rangle}.
\end{equation}
The Nusselt number can be equivalently expressed via various other spatial integrals that have the same value as \eqref{eq:NuDef} when averaged over infinite time.
One such alternative is the ratio of total transport at any fixed height $z \in [0, \pi]$, averaged horizontally and over time, to conductive transport, expressed as~\cite{Goluskin2015a}
\begin{equation} \label{eq:NuDef2}
{\rm{Nu}} = 1 + \frac{1}{\R} \left[ \overline{ \partial_z \langle\theta \rangle_x}(z) + \overline{ \langle \theta \partial_x \psi \rangle_x}(z) \right],
\end{equation}
where the horizontal average is denoted as
\begin{equation}
\langle f \rangle_x := \frac{1}{A\pi} \int_0^{A\pi} f(x) \, {\rm d}x.
\end{equation}
The correspondence between \eqref{eq:NuDef} and \eqref{eq:NuDef2} for truncated models derived by projecting the equations of motion onto a finite set of Fourier modes is preserved only by certain distinguished modal choices, including the modes we choose in the next section.

\section{Truncated model construction} \label{sec:ROM} 
Various ODE approximations of Rayleigh's PDE model have been derived as truncated Galerkin expansions, starting with the work of Saltzman~\cite{Saltzman62} and Lorenz~\cite{Lorenz63} in the 1960s. In these and subsequent studies of convection between stress-free boundaries, $\psi$ and $\theta$ are expanded in a Fourier basis in both directions. An ODE model is derived by selecting a finite set of modes, projecting $\psi$ and $\theta$ onto these modes, and projecting every term in the PDEs \eqref{eq:BE} onto the same modes. This yields a system of ODEs governing the amplitudes of the Fourier modes that are retained in the truncation. With sufficiently many modes included, the ODE dynamics are quantitatively close to the PDE dynamics, so integrating the ODE system amounts to direct numerical simulation of the PDEs with a spectral discretization of space. For various purposes, however, it is more useful to study a low-dimensional ODE model that differs quantitatively from the PDEs but captures certain qualitative features. The celebrated Lorenz equations~\cite{Lorenz63}, for instance, are a projection of Rayleigh's system onto only three modes.

In the present work we study an ODE model derived by projecting the Boussinesq equations onto the Fourier modes in the ansatz
\begin{equation} 
\label{eq:modes}
\begin{aligned} 
\psi(x,z,t) &= \psi_{11}(t) \sin (kx) \sin (z) + \psi_{12}(t) \cos (kx) \sin (2z) + \psi_{01}(t) \sin (z) + \psi_{03}(t) \sin (3z), \\[2pt]
\theta(x,z,t) &= \theta_{11}(t) \cos (kx) \sin (z) + \theta_{12}(t) \sin (kx) \sin (2z) + \theta_{02}(t) \sin (2z) + \theta_{04}(t) \sin (4z),
\end{aligned}
\end{equation}
where $k:=2/A$ is the fundamental horizontal wavenumber corresponding to a domain of aspect ratio $A$. The first and second subscripts on the mode amplitudes denote horizontal and vertical mode numbers, respectively. The chosen truncation includes the triplet $\{\psi_{11},\theta_{11},\theta_{02}\}$; a truncation with these three variables alone gives the Lorenz equations. It also includes the analogous triplet with vertical mode numbers doubled, $\{\psi_{12},\theta_{12},\theta_{04}\}$, alone yielding a rescaled version of the Lorenz equations. Modes in each triplet are coupled together by the two remaining modes, $\psi_{01}$ and~$\psi_{03}$. 

The modes included in the truncation \eqref{eq:modes} can capture flows whose horizontal velocities do not vanish after horizontal averaging. This is because the stream function modes $\psi_{01}$ and $\psi_{03}$ describe purely horizontal velocity fields. It was interest in such mean horizontal flows that motivated Howard and Krishnamurti~\cite{Howard86} to choose a truncation that is similar to \eqref{eq:modes} but omits the $\psi_{03}$ and $\theta_{04}$ modes, resulting in a six-dimensional ODE model. Their model helped illuminate a mean-flow instability but is not suitable for studying heat transport even as a low-order model because some of its trajectories are unbounded. Another drawback of their model is that expressions for time-averaged heat transport such as \eqref{eq:NuDef} and \eqref{eq:NuDef2} that are equivalent in the PDE dynamics give expressions that generally differ in the ODE dynamics once projected onto the chosen set of six modes. Thiffeault and Horton~\cite{Thiffeault1995, Thiffeault1996} found that adding the $\theta_{04}$ mode restores boundedness of trajectories and equality between the truncated versions of \eqref{eq:NuDef} and \eqref{eq:NuDef2}, as well as conservation of mechanical energy in the dissipationless limit.
Separately, Hermiz \emph{et al.}~\cite{Hermiz1995} found that adding the $\psi_{03}$ mode results in an ODE system whose solutions obey the truncated version of another PDE identity: $\partial_t \left\langle \nabla^2 \psi \right\rangle=0$, meaning that total vorticity is conserved.

Here we add both the $\theta_{04}$ and $\psi_{03}$ modes to the six chosen by Howard and Krishnamurti~\cite{Howard86} to construct an eight-dimensional truncation with all of the desirable properties mentioned above. We call the resulting system the HK8 model because it is the minimal extension of the six-mode model that restores these basic integral identities of the PDE. A version of the HK8 model was written down by Gluhovsky \emph{et al.}~\cite{Gluhovsky2002}, who confirmed that including the $\theta_{04}$ mode added by Thiffeault and Horton~\cite{Thiffeault1995, Thiffeault1996} and the $\psi_{03}$ mode added by Hermiz \emph{et al.}\ indeed combines the conservation properties of both. The HK8 model obtained by projecting the PDEs \eqref{eq:BE} onto the modes in \eqref{eq:modes} is~\cite{Goluskin2013}
\begin{equation} 
\label{eq:HK8}
\begin{aligned}
 \dot{\psi}_{11} &= -\sigma (k^2+1) \psi_{11} + \sigma \tfrac{k}{k^2+1} \theta_{11} { + \tfrac{k}{2} \tfrac{k^2+3}{k^2+1} \psi_{01} \psi_{12} }  {- \tfrac{3k}{2} \tfrac{k^2-5}{k^2+1} \psi_{12} \psi_{03} },\\
  { \dot{\psi}_{01} } & {  =-\sigma \, \psi_{01} - \tfrac{3k}{4} \psi_{11} \psi_{12} },\\
  { \dot{\psi}_{12}} & { = -\sigma (k^2+4) \psi_{12} - \sigma \tfrac{k}{k^2+4} \theta_{12} - \tfrac{1}{2} \tfrac{k^3}{k^2+4} \psi_{11} \psi_{01} } {\ + \tfrac{3k}{2} \tfrac{k^2-8}{k^2+4} \psi_{11} \psi_{03}}, \\
 \dot{\theta}_{11} &= -(k^2+1) \theta_{11} +  \mathcal{R} k \psi_{11} - k \psi_{11} \theta _{02}  { - \tfrac{k}{2} \psi_{01} \theta_{12} }  {\ +  \tfrac{3k}{2} \theta_{12} \psi_{03}}, \\
\dot{\theta}_{02} &= -4 \, \theta_{02} + \tfrac{k}{2} \psi_{11} \theta_{11}, \\
{ \dot{\theta}_{12} } & { = -(k^2+4) \theta_{12} - \mathcal{R} k \psi_{12} + \tfrac{k}{2} \psi_{01} \theta_{11} } {\ - \tfrac{3k}{2} \psi_{03} \theta_{11} + 2k \psi_{12} \theta_{04}}, \\
{\ \dot{\psi}_{03} } & {\ = -9\, \sigma \,  \psi_{03} + \tfrac{k}{4} \psi_{11} \psi_{12} },\\
{\ \dot{\theta}_{04} } & {\ = -16\,  \theta_{04} - k \psi_{12} \theta_{12} }.
\end{aligned}
\end{equation}
Discussions of the projection procedure for general truncations can be found elsewhere~\cite{Saltzman62, Thiffeault1995}.

The integral definitions of the Nusselt number for the PDE yield analogous expressions for the truncated model. We denote the truncated Nusselt number as $N$ to distinguish it from the PDE quantity Nu that it approximates. When applied to the modal expansions \eqref{eq:modes} that produce the HK8 truncation, the volume-averaged expression \eqref{eq:NuDef} for the Nusselt number becomes
\begin{equation}
\label{eq:Nu2}
N = 1+ \tfrac{k}{4\R} ( \ov{\psi_{11} \theta_{11} - \psi_{12} \theta_{12}} ),
\end{equation}
while the horizontally-averaged expression \eqref{eq:NuDef2} becomes
\begin{equation} 
\label{eq:Nu3}
N = 1 + \tfrac{1}{\R} (\ov{2 \theta_{02} + 4 \theta_{04}} ).
\end{equation}
It is shown in~\cite{Thiffeault1995} that the infinite-time averages \eqref{eq:Nu2} and \eqref{eq:Nu3} must be equal for all solutions of the HK8 model. Maximizing $N$ using either of the above expressions provides a finite-dimensional analogue of optimal heat transport for Rayleigh--B\'enard convection. For a given parameter set, the maximal $N$ is defined by
\begin{equation} \label{eq:maxN}
N^* := \sup_{\mathbf{x}(t)}{N},
\end{equation}
where the maximization is over all solutions $\mathbf{x}(t)$ of the HK8 model. The bifurcation structure of the HK8 model was explored in~\cite{Goluskin2013}, and upper bounds on heat transport that are not sharp in general were derived analytically in~\cite{Souza2015}. In~\S\ref{sec:particular} we explore particular steady states and time-dependent solutions that provide lower bounds on $N^*$, and in~\S\S\ref{sec:Numerical SDP}--\ref{sec:Quad} we establish upper bounds on $N^*$ using polynomial optimization. Combining the upper and lower bounds, we identify states that provide the maximal $N$ for the HK8 model and determine the regions in the $\sigma$--$\R$ plane where different types of solutions attain the maximal value $N^*$.

\section{Particular solutions of the HK8 model}
\label{sec:particular}
In this section we examine various particular solutions of the HK8 model, providing candidates for (and lower bounds on) the maximal $N$. We begin by summarizing the bifurcation structure of steady states of the HK8 model reported in~\cite{Goluskin2013} and verified here. Then we examine the heat transport along some time-dependent solutions. The maximum $N$ among these particular solutions provides a candidate for the supremum $N^*$ among all solutions, and we use it to judge the sharpness of upper bounds on $N^*$ reported in~\S\ref{sec:Numerical SDP} and~\S\ref{sec:Quad}. The value of $N$ among steady states is of particular importance as it is theorized that steady states maximize heat transport for Rayleigh--B\'enard convection \cite{Wen2020}.

\subsection{Steady states}
\label{sec:steady}

\begin{figure}[tp]
\begin{center}
\begin{tikzpicture}
\node at (0,0) {\includegraphics[trim = 0 10 0 0, clip]{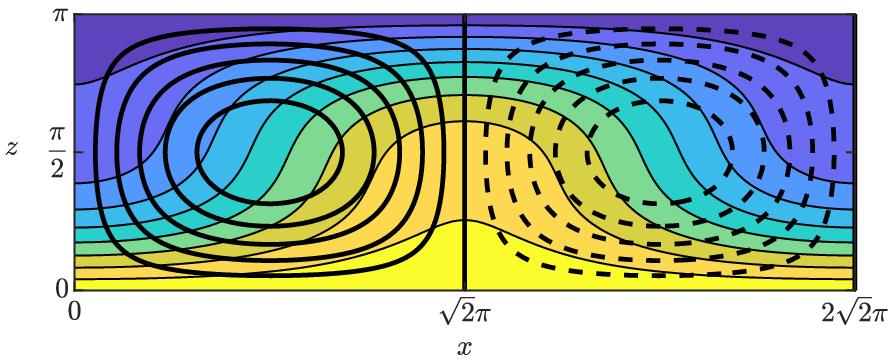}};
\node at (-5.1,1.33) {\small (a)};
\end{tikzpicture}
\begin{tikzpicture}
\node at (0,0) {\includegraphics[trim = 0 10 0 0, clip]{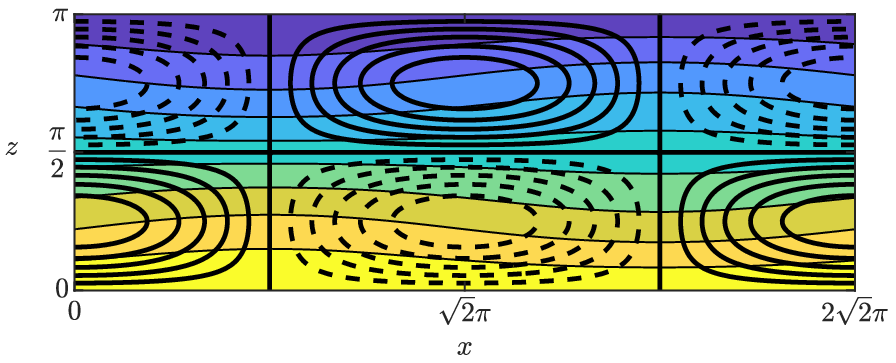}};
\node at (-5.1,1.32) {\small (b)};
\end{tikzpicture}
\begin{tikzpicture}
\node at (0,0) {\includegraphics[trim = 0 0 0 0, clip]{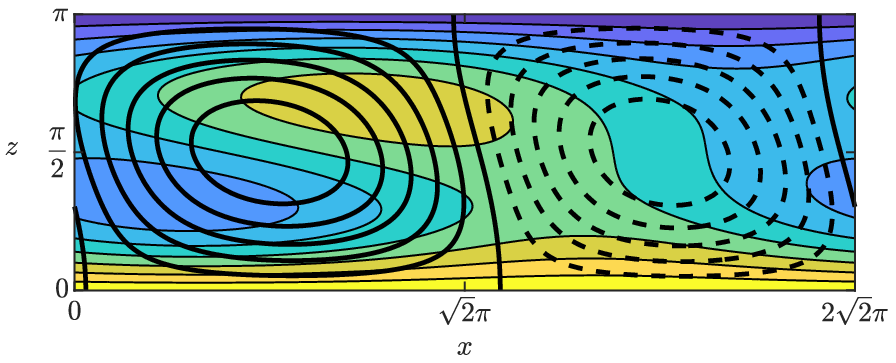}};
\node at (-5.1,1.52) {\small (c)};
\end{tikzpicture}
\end{center}
\caption{Streamlines overlaid on contours of temperature ($T$) for approximations of steady convection states whose mode amplitudes in the truncated Galerkin expansion \eqref{eq:modes} are equilibria of the HK8 model with ${(k^2,\sigma) = (1/2,10)}$. Each of the three types of equilibria is depicted near its onset: (a) an $L_1$ state at $\R = 10$, (b) an $L_2$ state at $\R = 185$, and (c) a $TC$ state at $\R = 150$.  The $T$ scale ranges from 0 (dark) to 1 (light). Positive and negative vorticity is indicated by solid and dashed streamlines, respectively. The $TC$ states, in particular, display unphysical behavior due to the truncation of the PDE, evidenced by the internal temperature maximum in (c).}
\label{fig:Stream}
\end{figure}

At sufficiently small $\R$ the zero equilibrium is globally attracting. This solution corresponds to the purely conductive state in the PDE. The HK8 system has three branches of nonzero equilibria that we call $L_1$, $L_2$, and $TC$ in analogy with~\cite{Howard86}. At the Rayleigh number $\R_{L_1}$, defined by
\begin{equation} \label{eq:RL1}
\R_{L_1} := \frac{(k^2+1)^3}{k^2},
\end{equation}
the zero state undergoes a pitchfork bifurcation giving rise to $L_1$ equilibria that exist for all $\R>\R_{L_1}$, so-named because the only nonzero modes are the first Lorenz triplet,
\begin{equation} \label{eq:L1}
\psi_{11} = \pm \sqrt{8} \, \tfrac{1}{k^2+1} \sqrt{\R - \R_{L_1}}, \quad \theta_{11} = \pm \sqrt{8} \, \tfrac{k^2+1}{k} \sqrt{\R - \R_{L_1}}, \quad \theta_{02} = \R - \R_{L_1}.
\end{equation}
As shown in Figure~\ref{fig:Stream}(a), the $L_1$ states are an approximation of a PDE steady state with a pair of counter-rotating convection rolls. The Rayleigh number $\R_{L_1}$ reaches a minimum of $27/4$ when $k^2 = 1/2$, corresponding exactly to the onset of convection for 2D stress-free Rayleigh--B\'enard convection. Therefore, we define the critical Rayleigh number $\R_c$ as
\begin{equation} \label{eq:Rc}
\R_c := \frac{27}{4}.
\end{equation} 
At the Rayleigh number $\R_{L_2}$, given by
\begin{equation} \label{eq:RL2}
\R_{L_2} := \frac{(k^2+4)^3}{k^2},
\end{equation}
the zero solution undergoes a second pitchfork bifurcation, giving rise to $L_2$ equilibria when $\linebreak \R>\R_{L_2}$. There the nonzero modes are the second Lorenz triplet,\begin{equation} \label{eq:L2}
\psi_{12} = \pm \sqrt{8} \, \tfrac{1}{k^2+4} \sqrt{\R - \R_{L_2}}, \quad \theta_{12} = \mp \sqrt{8} \, \tfrac{k^2+4}{k} \sqrt{\R - \R_{L_2}}, \quad \theta_{04} = \tfrac{1}{2}(\R - \R_{L_2}).
\end{equation}
As shown in Figure~\ref{fig:Stream}(b), the $L_2$ states are an approximation of a PDE steady state with a two-by-two array of convection rolls. In the full PDE there are counterparts to the $L_1$ and $L_2$ branches that bifurcate from the conductive state at the same $\R$ values. They agree asymptotically with the truncated states in the weakly nonlinear regime, but this resemblance decreases as $\R$ grows.

The $L_1$ and $L_2$ equilibria are scaled versions of the nonzero equilibria of the Lorenz equations. In fact, if the HK8 equations are restricted to the three-dimensional subspaces spanned by the nonzero variables in either \eqref{eq:L1} or \eqref{eq:L2}, the resulting dynamics are equivalent to those of the Lorenz equations. Thus, for any solution to the Lorenz
equations, a corresponding solution to the HK8 system can be obtained by a suitable linear
change of variables. However, the dynamics of the HK8 model on these lower-dimensional manifolds appears to be unstable to off-manifold perturbations for sufficiently large $\R$.

\begin{figure}[tp]
	\begin{center}
	\begin{tikzpicture}[>=stealth, line width=.7pt]
		\node[anchor=north west, inner sep=0] at (0,0) {\includegraphics[scale=.5]{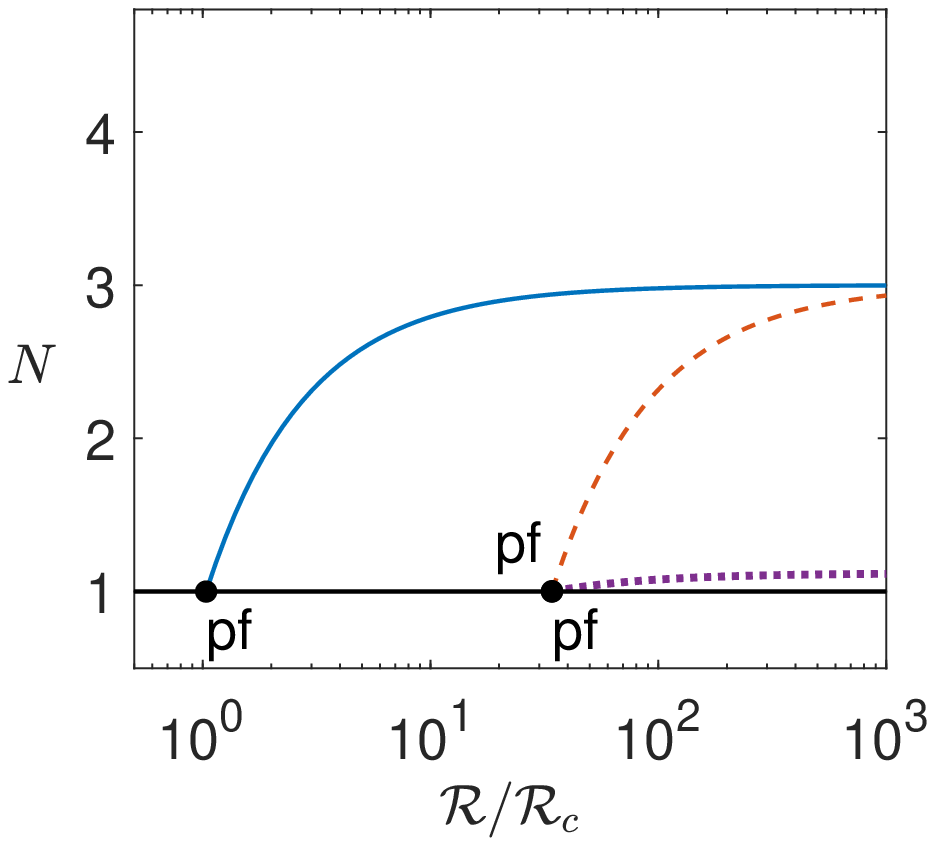}};
		\node[scale=1] at (1.0,-.4) {\large I};
		\node[scale=1] (A) at (3.4,-.56) {$L_1$};
		\node (B) at (2.8,-1.6) {};
		\draw[->] (A) -- (B);
		\node[scale=1] (C) at (2.19,-1.9) {$L_2$};
		\node (D) at (3.14,-2.5) {};
		\draw[->] (C) -- (D);
		\node[scale=1] (E) at (4.1,-2) {$TC$};
		\node (F) at (3.5,-3.04) {};
		\draw[->] (E) -- (F);
		\end{tikzpicture}
		\begin{tikzpicture}
		\node[anchor=north west, inner sep=0] at (0,0) {\includegraphics[scale=.5, trim = 9 0 0 0, clip]{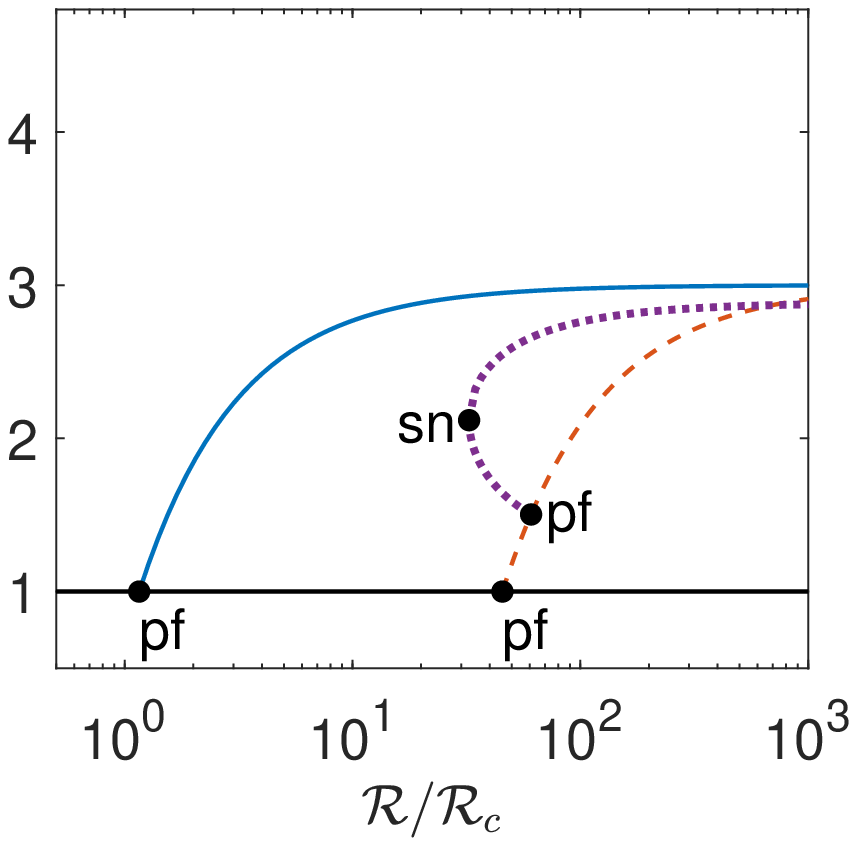}};
		\node[scale=1] at (.44,-.4) {\large II};
		\end{tikzpicture}
		\begin{tikzpicture}
		\node[anchor=north west, inner sep=0] at (0,0) {\includegraphics[scale=.5, trim= 9 0 0 0, clip]{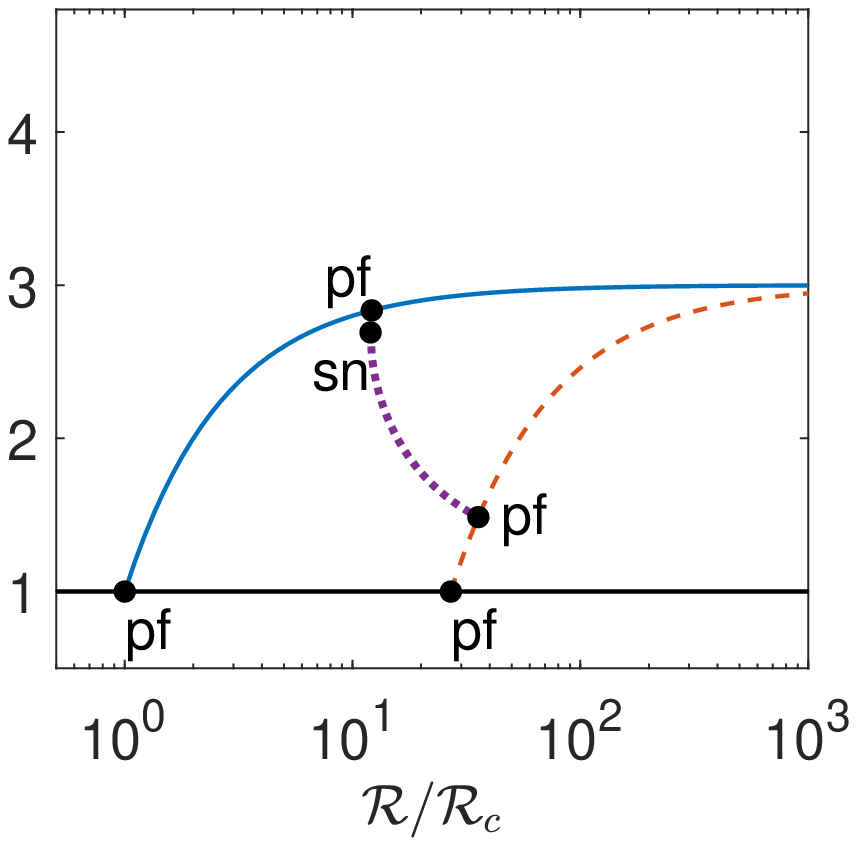}};
		\node[scale=1] at (.48,-.4) {\large III};
		\end{tikzpicture} \\[4pt]
		\begin{tikzpicture}
		\node[anchor=north west, inner sep=0] at (0,0) {\includegraphics[scale=.5]{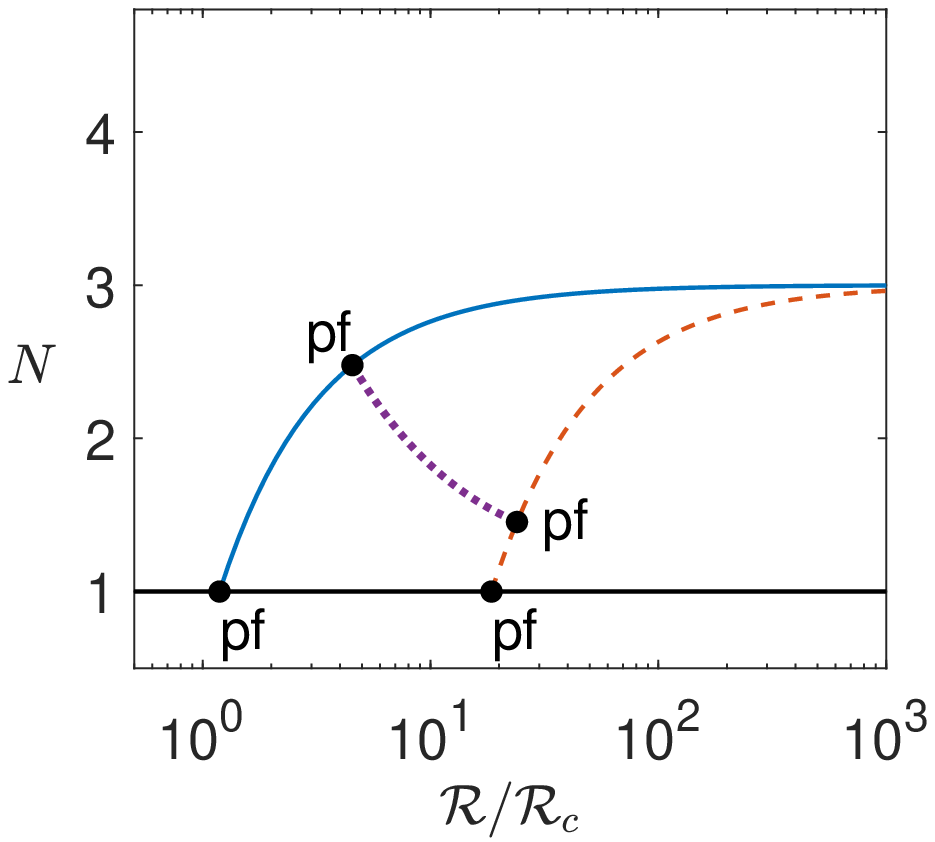}};
		\node[scale=1] at (1.02,-.4) {\large IV};
		\end{tikzpicture}
		\begin{tikzpicture}
		\node[anchor=north west, inner sep=0] at (0,0) {\includegraphics[scale=.5, trim = 9 0 0 0, clip]{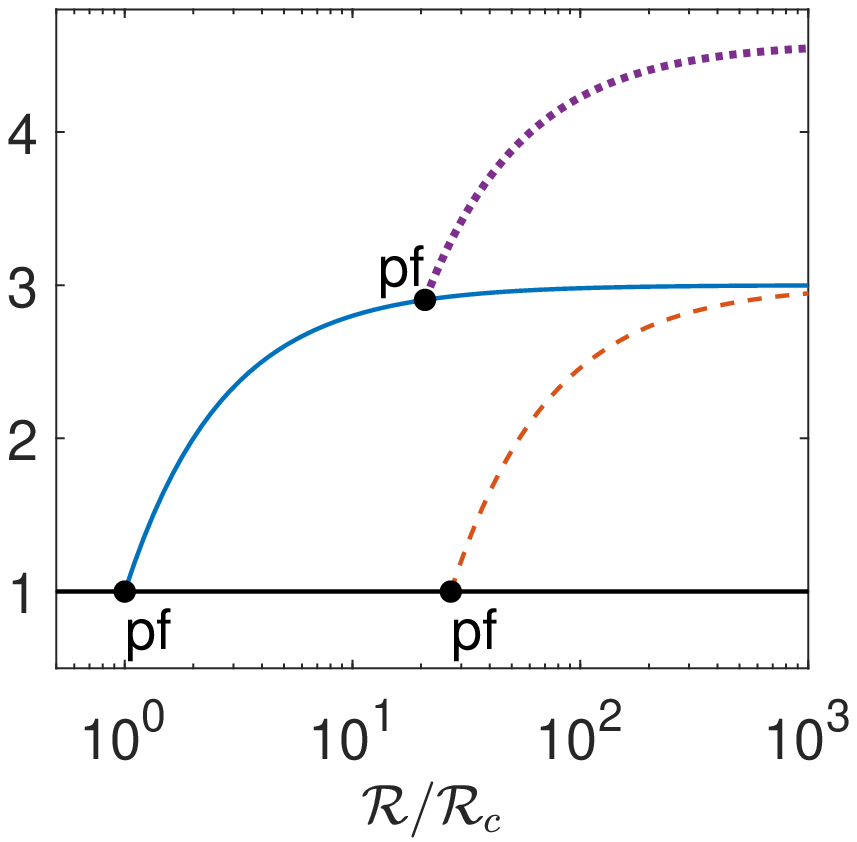}};
		\node[scale=1] at (.44,-.4) {\large V};
		\end{tikzpicture}
				\begin{tikzpicture}[>=stealth, line width=1pt]
				\node[anchor=south west, inner sep=0] at (0,0) {\includegraphics{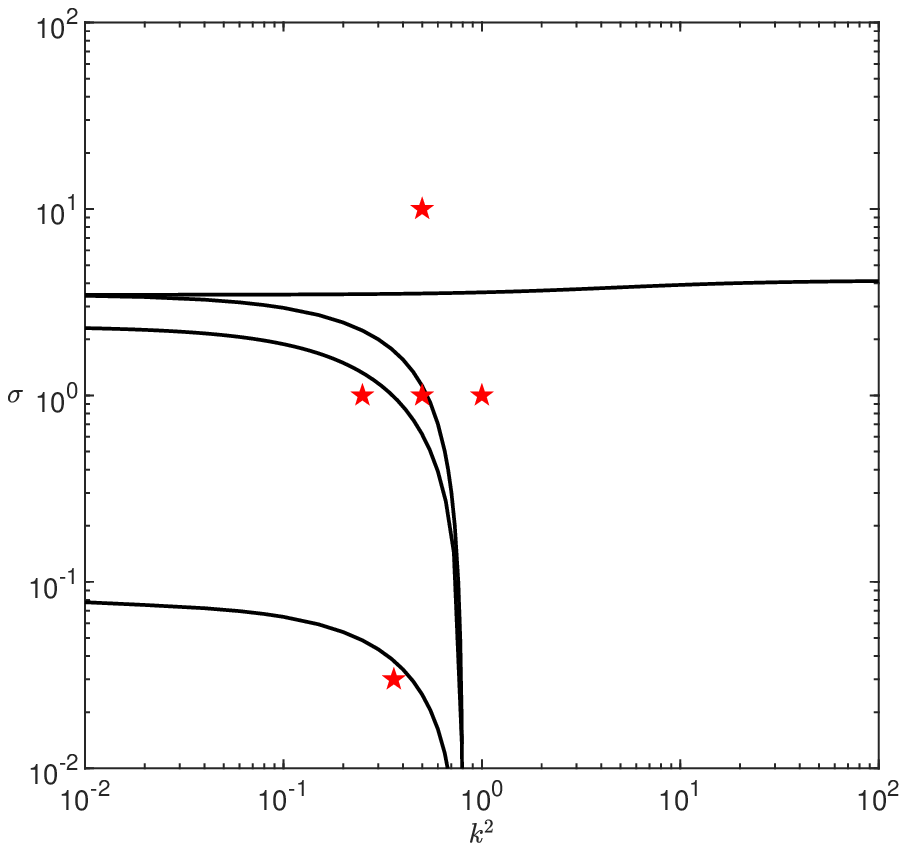}};
				\node[scale=1] at (2.7,1.55) {\large I};
				\node[scale=1] at (2.7,3.7) {\large II};
				\node[scale=1] (3a) at (2.7,6.6) {\large III};
				\node (3b) at (2.7,5.12) {};
				\draw[->] (3a) -- (3b);
				\node[scale=1] at (6.7,3.7) {\large IV};
				\node[scale=1] at (4.8,7.1) {\large V};
				\end{tikzpicture}
	\end{center}
\caption{Examples of the five bifurcation structures of steady states of the HK8 model, with $\R$ as the bifurcation parameter (top). The $k^2$--$\sigma$ parameter regimes where each bifurcation structure occurs are also shown and numbered correspondingly (bottom). Stars in parameter space ({\red $\bigstar$}) indicate the particular values for each example bifurcation diagram above. Pitchfork (`pf') and saddle-node (`sn') bifurcations are labeled. Stability of steady states and locations of Hopf bifurcations are not indicated. The results in this figure were reported by~\cite{Goluskin2013} and independently verified here. In the region I diagram, the $TC$ branch bifurcates from the $L_2$ branch at a Rayleigh number very close to $\R_{L_2}$.}
\label{fig:bif}
\end{figure}

The third type of equilibria found in the HK8 model are called the $TC$ states because they correspond to so-called tilted cells~\cite{Howard86}. As shown in Figure~\ref{fig:Stream}(c), a pair of steady convection rolls produces a mean horizontal flow whose direction breaks the symmetry of the $L_1$ and $L_2$ states. All eight modes are nonzero in the $TC$ states, and here we compute them numerically using the numerical continuation software MATCONT~\cite{Dhooge2006}. The temperature field in Figure~\ref{fig:Stream}(c) is somewhat unphysical, as evidenced by the internal maxima of the steady temperature field, reflecting the fact that the truncated model is not capturing the full PDE dynamics at this $\R$.

Depending on the fixed values of $k^2$ and $\sigma$, as $\R$ is varied there are five possible bifurcation structures where the $TC$ branches connect to the $L_1$ or $L_2$ branches, or both. Figure~\ref{fig:bif} shows an example of each possible bifurcation structure, along with the regimes in the $k^2$--$\sigma$ parameter plane where each structure occurs. In regimes III--V, a pair of $TC$ branches connects to each $L_1$ branch in a pitchfork bifurcation at $\R_{TC_1}$, where~\cite{Goluskin2013}
\begin{equation} 
\label{eq:Rs}
\frac{\R_{TC_1}}{\R_{L_1}} = 1 + \frac{27 \sigma^2}{k^2+1} \frac{k^4 + 5 k^2 + 7}{(10 \sigma + 3 \sigma^2) (k^2 + 1)^2 + 2(k^2+4)(5 k^2 - 4)}.
\end{equation}
The above denominator is negative in regimes I and II, so the $TC$ and $L_1$ branches do not connect. Similarly, in regimes I--IV, a pair of $TC$ branches connects to each $L_2$ branch in a pitchfork bifurcation at $\R_{TC_2}$, where
\begin{equation} \label{eq:Rs2}
\frac{\R_{TC_2}}{\R_{L_2}} = 1 + \frac{27 \sigma^2}{k^2+4} \frac{k^4 + 5 k^2 + 7}{(10 \sigma - 3 \sigma^2) (k^2 + 4)^2 + 2(k^2+1)(5 k^2 +11)}.
\end{equation}
For parameter combinations in regime V the $TC$ and $L_2$ branches do not connect, as the denominator of \eqref{eq:Rs2} is negative. Counterparts to the $TC$ branches have been observed for the full PDE, at least for some values of $k^2$ and $\sigma$~\cite{Goluskin2013}. The bifurcations connecting the $TC$ branches to the $L_1$ and $L_2$ branches are quantitatively accurate only in the $\sigma\to0$ limit since this is when they occur in the weakly nonlinear regime. 

The $L_1$, $L_2$, and $TC$ states are the only nonzero steady states of the HK8 model \cite{Goluskin2013}, so finding the maximum $N$ among them at a given parameter set yields the maximum heat transport by any steady state. Evaluating \eqref{eq:Nu2} or \eqref{eq:Nu3} to find $N$ in the HK8 model gives the heat transport by the $L_1$ and $L_2$ equilibria:
\begin{align} 
\label{eq:NL12}
N_{L_1} &= 3 -  2 \, \frac{\R_{L_1}}{\R}, &
N_{L_2} &= 3 -  2 \, \frac{\R_{L_2}}{\R}.
\end{align}
Both values approach 3 as $\R\to\infty$, but $N_{L_1} > N_{L_2}$ at any parameters where both states exist. We computed $N_{TC}$ numerically at many parameter values in all five parameter regimes. In regimes I--IV we found $N_{L_1}>N_{TC}$ in all cases, meaning the $L_1$ branch maximizes heat transport among steady states. In regime V, at sufficiently large $\R$ the $TC$ branch maximizes $N$ among steady states. Whether these maximal steady $N$ values are also maximal among time-dependent solutions remains to be determined by the bounds computed in~\S\ref{sec:Numerical SDP} and~\S\ref{sec:Quad}. We note that the results at large $\R$ are unlikely to be representative of the PDE: the values of $N_{L_1}$ begin to deviate from the values of Nu for the analogous steady solutions of the Boussinesq equations (i.e., the primary branch of convection rolls that arises as the first instability of the conduction state) near $\R = 2\R_c$. Mean horizontal flow, exhibited in the HK8 system by the $TC$ equilibria and various time-dependent solutions, has been observed to reduce heat transport in simulations of 2D Rayleigh–-B\'enard convection in a horizontally periodic domain with stress-free boundaries \cite{GJFS2014JFM}.

All three nonzero steady branches can undergo Hopf bifurcations. Determining $k^2$ and $\sigma$ for the various pitchfork, saddle-node, and Hopf bifurcations to exist (with $\R$ as the bifurcation parameter) subdivides the five parameter regimes of Figure~\ref{fig:bif} into 16 regimes, as reported in~\cite{Goluskin2013}. The Hopf bifurcation on the $L_1$ branch involves only the three Lorenz modes $\{\psi_{11},\theta_{11},\theta_{02}\}$ and is precisely the Hopf bifurcation found in the Lorenz equations~\cite{Lorenz63}. With the present variables (scaled differently from the Lorenz equations), the Hopf bifurcation exists when $\sigma > 1 + 4/(k^2+1)$ and occurs at $\R_{H_1}$, where
\begin{equation} \label{eq:Hopf1}
\frac{\R_{H_1}}{\R_{L_1}} = 1 + (\sigma + 1)\frac{\sigma(k^2+1) + (k^2+5)}{\sigma(k^2+1) - (k^2+5)}.
\end{equation}
Since the $L_2$ subspace is another rescaling of the Lorenz equations, its Hopf bifurcation is similar. It exists when $\sigma > 1 + 16/(k^2+4)$ and occurs at $\R_{H_2}$, where
\begin{equation} \label{eq:Hopf2}
\frac{\R_{H_2}}{\R_{L_2}} = 1 + (\sigma + 1)\frac{\sigma(k^2+4) + (k^2+20)}{\sigma(k^2+4) - (k^2+20)}.
\end{equation}
Additional Hopf bifurcations may occur on the $TC$ branch, as detailed in~\cite{Goluskin2013}.

Periodic states emerging from Hopf bifurcations of the $L_1$ and $L_2$ branches remain in their respective subspaces of Lorenz triplets, where the dynamics are equivalent to those of the Lorenz equations. In the Lorenz equations, the truncated Nusselt number is maximized at the nonzero equilibria~\cite{Goluskin2018, Malkus1972} that correspond to the $L_1$ equilibria. As a result, these three-dimensional periodic orbits cannot produce larger heat transport than $N_{L_1}$. It remains possible that time-dependent states involving all eight modes can have larger $N$ than all steady states; we examine such solutions in the next subsection. 

\subsection{Time-dependent states}
\label{sec:Direct}

When time-dependent trajectories are not known exactly, time averages may be estimated from \eqref{eq:timeAvg} by numerically integrating the system starting from particular initial conditions and averaging over sufficiently large time intervals. It is generally not possible to compute the supremum \eqref{eq:maxN} of $N$ directly in this way since the number of possible initial conditions is infinite, and the dependence of $N$ on initial conditions is non-convex. As in the previous subsection, the goal of directly computing time averages is to identify candidates for the maximal heat transport, and to compare the resulting values of $N$ to the upper bounds in \S\S\ref{sec:Numerical SDP}--\ref{sec:Quad}.
%Convergence might be obtained through direct application of \eqref{eq:timeAvg}, but this may be prohibitively costly to compute accurately: for example, it may be particularly difficult to compute the average along slowly converging or unstable solutions.
%Furthermore, it is generally not possible to explore trajectories emanating from all possible initial conditions.

We numerically integrated \eqref{eq:HK8} to search for attracting time-dependent solutions of the HK8 model with $(k^2,\sigma)=(1/2,10)$ fixed. These parameter values lie in regime V of Figure~\ref{fig:bif}, and they correspond to the standard choice $(\beta,\sigma)=(8/3,10)$ in the Lorenz equations. This $k$ value minimizes the Rayleigh number $\R_{L_1}$ of convective instability in both the HK8 model and the PDE. Numerical integration was carried out using MATLAB's \texttt{ode45} function with absolute and relative tolerances of $10^{-12}$ and $10^{-9}$, respectively, and all other settings at their default values.
% Some unstable periodic orbits were computed by numerically continuing them from Hopf bifurcations using the software MATCONT. 
The time-averaged Nusselt number \eqref{eq:Nu3} was computed by averaging periodic trajectories over several full periods and by averaging aperiodic trajectories over $10^4$ to $10^5$ time units after initial transients.

\begin{figure}[tp]
\begin{center}
\vspace{.1cm}
\includegraphics{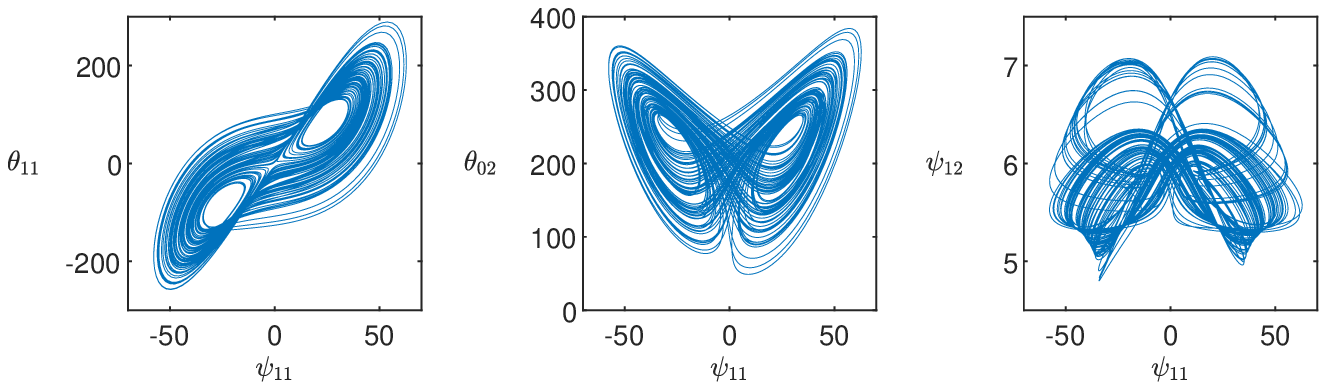} \\[4pt]
\includegraphics{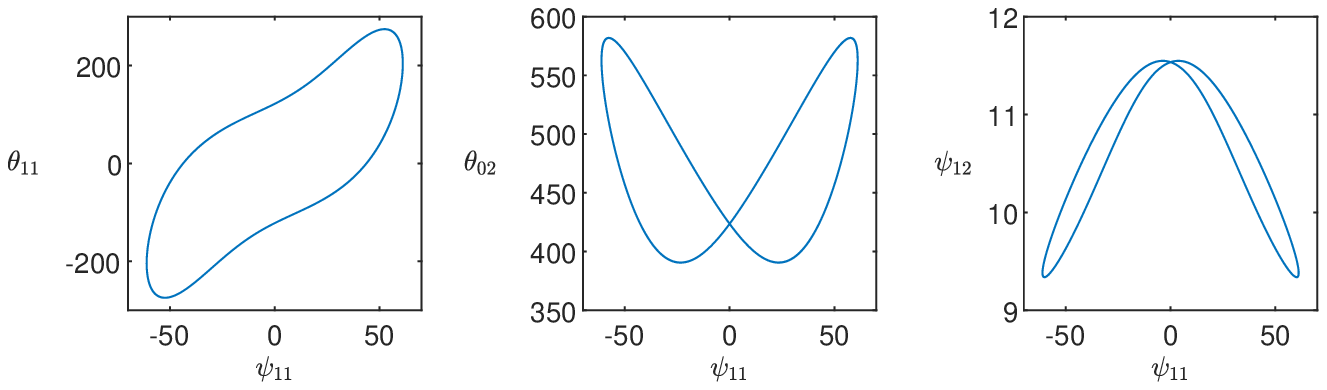}
\caption{\label{fig:phase} Evolution of the Lorenz modes $(\psi_{11},\theta_{11},\theta_{02})$ and the $\psi_{12}$ mode in projections of phase space for trajectories of the HK8 system. The top row displays orbits of an apparently chaotic trajectory at $\R=250$, while the bottom row depicts a stable periodic trajectory at $\R=500$. In each case, all eight variables are generically nonzero along orbits.}
\end{center}
\end{figure}

When $(k^2,\sigma)=\nolinebreak(1/2,10)$, the $TC$ branch has subcritical Hopf bifurcations at $\R \approx 21.8 \, \R_c$ and $\R \approx 999 \, \R_c$. Above the first bifurcation, numerical integration with a variety of randomly generated initial conditions gives trajectories where all eight modes appear to be chaotic. The top row of figure~\ref{fig:phase} shows part of such a trajectory at $\R = 250$. We were able to find such seemingly chaotic trajectories for $\R/\R_c \in (21.8,45)$ and again at $\R\gtrsim 1.8 \times 10^3 \, \R_c$; between these two intervals, the only states we found using numerical integration are periodic. An example at $\R=500$ is shown in the bottom row of Figure~\ref{fig:phase}. Bistability between periodic and seemingly chaotic states was found for $\R \gtrsim 1.8 \times 10^3 \, \R_c$, where some initial conditions produced solutions that approached periodic trajectories, while others tended towards a nonperiodic attractor similar to that depicted in the top row of Figure \ref{fig:phase}. Bistable behavior was not identified for any smaller values of $\R$. The system exhibits period doubling bifurcations as $\R$ is increased from the Hopf bifurcation of the $TC$ branch and as $\R$ is decreased from $\R \approx 46 \, \R_c$ into the nonperiodic regime, but the possibility of a full period doubling cascade was not explored in detail.

\begin{figure}[tp]
\begin{center}
\begin{tikzpicture}[>=stealth,line width=.8pt]
\node[anchor=south west, inner sep=0] at (0,0) {\includegraphics[scale=.5]{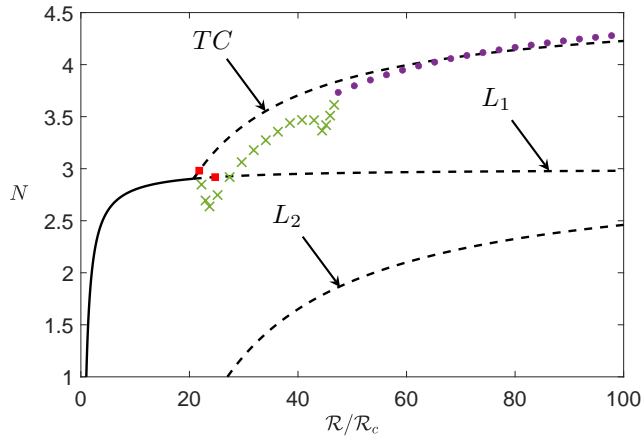}};
\node[scale=1] (A) at (6.49,4.5) {$L_1$};
\node (B) at (7.29,3.42) {};
\draw[->] (A) -- (B);
\node[scale=1] (C) at (3.71,2.95) {$L_2$};
\node (D) at (4.51,1.87) {};
\draw[->] (C) -- (D);
\node[scale=1] (E) at (2.72,5.29) {$TC$};
\node (F) at (3.52,4.21) {};
\draw[->] (E) -- (F);
\end{tikzpicture}
\end{center}
\caption{Nusselt numbers of steady and time-dependent solutions of the HK8 model for $(k^2,\sigma) = (1/2,10)$. Symbols denote time averages over time-dependent states that are periodic ({\color{purpdot}$\bullet$}) or appear to be chaotic ({\color{greenx}$\times$}).
Solid lines denote linearly stable equilibria while dashed lines denote unstable equilibria. $TC$ equilibria are stable only in the small interval of $\R$ between their emergence from the $L_1$ branch and the subsequent Hopf bifurcation. Locations of Hopf bifurcations are indicated by red squares ({\red \tiny $\blacksquare$}). }
\label{fig:N_eq}
\end{figure}

Figure~\ref{fig:N_eq} shows the values of $N$ versus $\R$ for all steady states and time-dependent states found using time integration. The $N$ values in the nonperiodic regime all lie below the steady state maximum. As $\R$ is raised, the $N$ values of the periodic states surpasses $N_{TC}$, meaning that heat transport is \emph{not} maximized by a steady state at large $\R$. At such large $\R$, however, the HK8 model is not expected to closely reflect behavior of the PDE. At smaller $\R$ we did not find any time-dependent states with $N$ larger than the steady state maximum. For $\R \lesssim 71 \, \R_c$ the steady states indeed maximize heat transport, as follows from our sharp upper bounds on $N^*$ in~\S\ref{sec:Numerical SDP} that are equal to $\max\{N_{L_1},N_{TC}\}$.

\section{Bounding time averages using sum-of-squares polynomial optimization} \label{sec:SOS}
To bound the optimal time-averaged heat transport among all trajectories in the HK8 model---the supremum \eqref{eq:maxN} of $N$ as defined equivalently by \eqref{eq:Nu2} or \eqref{eq:Nu3}---we use a general method for bounding infinite-time averages in ODEs and PDEs.  As explained below, the method relies on constructing \emph{auxiliary functions} that satisfy certain inequalities, implying a bound on the time average of interest. In the case of ODEs with polynomial right-hand sides, bounds on averages of polynomial quantities can be sought using polynomial auxiliary functions. In such cases the conditions on auxiliary functions amount to nonnegativity constraints on polynomial expressions, and auxiliary functions can be constructed computationally using methods of polynomial optimization with sum-of-squares (SOS) constraints. The SOS approach was suggested only a few years ago~\cite{Cherny2014} and has surpassed other methods for the few systems to which it has been applied~\cite{Fantuzzi2016, Goluskin2018, Goluskin2019}. This approach is described for general ODEs in the present section. Numerical and analytical results from its application to bounding time-averaged heat transport in the HK8 model appear in~\S\ref{sec:Numerical SDP} and~\S\ref{sec:Quad}, respectively.

Consider a well-posed autonomous ODE ${\dot{\mathbf{x}} = \mathbf{f}(\mathbf{x})}$, where ${\mathbf{f}: \mathbb{R}^n \to \mathbb{R}^n}$ is continuously differentiable and each trajectory ${\mathbf{x}(t) \in \mathbb{R}^n}$ with initial condition $\bx(0) = \bx_0$ remains bounded forward in time. Let $\Phi: \mathbb R^n\to\mathbb R$ be a continuous quantity of interest whose infinite-time average \eqref{eq:timeAvg} along the trajectory $\bx(t)$ emanating from $\bx_0$ is denoted by $\ov{\Phi} (\bx_0)$. Define the supremum of the time average among all trajectories as
\begin{equation}
\label{eq:Phi*}
\ov\Phi^* := \sup_{\bx_0 \in \mathbb{R}^n} \ov{\Phi} (\bx_0).
\end{equation}
Our aim is to seek an upper bound $\ov{\Phi}^* \leq U$ that applies uniformly to all trajectories. In our application to the HK8 model in the following sections, we choose $\Phi= 1+\tfrac{1}{\R} \left(2\theta_{02}+4\theta_{04}\right)$ because then $\overline{\Phi} = N$, and upper bounds apply to the supremum \eqref{eq:maxN} of $N$ over all trajectories.

To construct global upper bounds on $\ov{\Phi}^*$, introduce an auxiliary function ${V:\mathbb R^n\to\mathbb R}$ in the class $C^1$ of continuously differentiable functions. Any such $V$ remains bounded along bounded trajectories. This implies ${\overline{\mathbf f\cdot\nabla V}=0}$ on every trajectory, where the gradient is with respect to $\mathbf x$, since
\begin{equation}
\label{eq:fGradV}
\overline{\mathbf f(\mathbf x(t))\cdot\nabla V(\mathbf x(t))} =
\ov{\tfrac{\rm d}{{\rm d}t} V(\mathbf{x}(t))} = \lim_{\tau \to \infty} \frac{1}{\tau} \Big[V(\mathbf x(\tau)) - V(\mathbf x(0))\Big] = 0.
\end{equation}
To produce upper bounds on $\overline\Phi$, use the identity \eqref{eq:fGradV} to estimate
\begin{equation}
\label{eq: bound1}
\overline\Phi = \overline{\Phi+\mathbf f\cdot \nabla V}\le
\sup_{\mathbf x\in \mathbb{R}^n}\left[\Phi(\bx)+\mathbf f(\bx)\cdot\nabla V(\bx)\right].
\end{equation}
This is useful because computing or estimating the right-hand supremum requires no knowledge of trajectories. While $\Phi(\bx)$ may be unbounded over $\mathbb R^n$, a judicious choice of $V(\bx)$ makes the above supremum finite. Since \eqref{eq: bound1} applies to all trajectories in bounded systems and for all $V\in C^1$, it remains true when we maximize $\ov\Phi$ over initial conditions and minimize the upper bound over $V$ to find
\begin{equation}
\label{eq: bound2}
\ov\Phi^* \le \inf_{V\in C^1}
\sup_{\mathbf x\in \mathbb{R}^n}\left[\Phi(\bx)+\mathbf f(\bx)\cdot\nabla V(\bx)\right].
\end{equation}
An equivalent way to express this inequality is
\begin{equation}
\label{eq: bound3}
\ov\Phi^* \leq \inf_{\substack{V\in C^1\\[2pt]S\ge0}} U,
\end{equation}
where $S\ge0$ indicates the pointwise nonnegativity on $\mathbb{R}^n$ of the function
\begin{equation}
\label{eq:S}
S(\bx) := U -  \Phi(\bx) - \mathbf{f}(\bx) \cdot \nabla V(\bx).
\end{equation}
In fact, for all bounded well-posed ODEs and continuous $\Phi(\bx)$, it has been proved that the inequality in \eqref{eq: bound2} is an equality~\cite{Tobasco2018} if the maximization is taken over a compact domain containing the attracting region of the ODE. The practical challenge is to construct an auxiliary function $V$ such that $S\ge0$ can be verified with the smallest possible upper bound $U$.

The right-hand side of \eqref{eq: bound3} is an optimization problem over the infinite-dimensional space $C^1$. Letting $V$ be a polynomial of degree no larger than $d$ gives an optimization problem over the finite-dimensional vector space $\mathbb{P}_{n,d}$ of such polynomials in $n$ variables. This bound need not be sharp for finite $d$, but it is proven to converge to $\ov{\Phi}^*$ as $d\to\infty$ for any dynamical system where trajectories remain in a compact set forward in time \cite{Korda2018,Lakshmi2020}. The resulting optimization problem is finite-dimensional and convex in $V$, but still it is not tractable since deciding nonnegativity of the polynomial $S$ is NP-hard in general. We thus use a relaxation that has become standard for polynomial optimization since its introduction two decades ago~\cite{Lasserre2001, Nesterov2000, Parrilo2000}: nonnegativity of $S$ over $\mathbb{R}^n$ is ensured by the stronger requirement that $S$ admits a representation as a sum of squares of other polynomials. That is, we require $S$ to lie in the set $\Sigma_n$ of SOS-representable polynomials in $n$ variables. If $V$ has fixed maximum degree $d$, the upper bound from the resulting polynomial optimization problem is~\cite{Cherny2014, Fantuzzi2016, Goluskin2018}
\begin{equation}
\label{eq:Opt}
\ov\Phi^* \leq U^*_d := \inf_{V\in \mathbb{P}_{n,d}}~U \quad
\rm{s.t.} \quad S\in\Sigma_n.
\end{equation}
The SOS-constrained polynomial optimization problem on the right-hand side of \eqref{eq:Opt} is computationally tractable if $d$ and $n$ are not too large. The ODE studied here has dimension $n=8$, and computations with $d\le6$ run in seconds on a laptop. Convergence of the upper bound to $\ov{\Phi}^*$ as $d \to \infty$ is not guaranteed by the theorems of \cite{Korda2018,Lakshmi2020} because they use slightly different SOS conditions implying nonnegativity only on a compact set. However, in practice the bounds \eqref{eq:Opt} often converge rapidly to $\ov{\Phi^*}$~\cite{Fantuzzi2016,Goluskin2018,Goluskin2019}.

The usual computational approach to solving an SOS-constrained optimization problem as in \eqref{eq:Opt} is to reformulate it as a semidefinite program (SDP), a standard type of conic optimization problem. This is done by representing the polynomial $S$ using a symmetric Gram matrix $\Q$ by
\begin{equation}
\label{eq: Gram}
S = \mathbf{b}^\mathsf{T} \Q \mathbf{b},
\end{equation}
where $\mathbf b(\bx)$ is a vector of polynomial basis functions. The vector $\mathbf b$ is chosen such that $S$ is in the span of the scalar polynomial entries in $\mathbf b \mathbf b^\mathsf{T}$, so that at least one $\Q$ exists satisfying \eqref{eq: Gram}. Furthermore, $S$ is an SOS polynomial if and only if at least one $\Q$ satisfying \eqref{eq: Gram} is positive semidefinite~\cite{Powers1998}. For a chosen basis $\bf b$, the polynomial optimization \eqref{eq:Opt} can be formulated equivalently as
\begin{equation}
\label{eq:SDP}
\ov\Phi^* \leq U^*_d := \min_{V\in \mathbb{P}_{n,d}}~U \quad \rm{s.t.} \quad
\begin{array}[t]{l}
S  = \mathbf{b}^\mathsf{T} \Q \mathbf{b}, \\
\Q\succeq 0.
\end{array}
\end{equation}
In the above optimization problem, the bound $U$ and the coefficients of the polynomial $V$ are tunable. The equality $S =\mathbf{b}^\mathsf{T} \Q \mathbf{b}$ is enforced by expanding out the right-hand product and matching coefficients on each monomial term, amounting to affine constraints on the entries of $\Q$. 
Thus, the optimization is over symmetric matrices $\Q$ subject to affine and semidefinite constraints that depend linearly on the tunable variable $U$ and the coefficients of the ansatz for $V$. These two types of constraints on a semidefinite matrix are what define an SDP~\cite{Boyd2004}. Various software is available to solve SDPs computationally, and we report numerical results in the next section. Analytical solutions are possible in cases leading to very small SDPs, as in~\cite{Goluskin2018, Powers1998}, and we report some analytical results in~\S\ref{sec:Quad}.

\section{Numerical upper bounds} 
\label{sec:Numerical SDP}
To compute upper bounds on $N^*$---the maximum of $N$ among all trajectories in the HK8 model---we numerically solved polynomial optimization problems of the form \eqref{eq:SDP}. In the definition \eqref{eq:S} of $S$, the vector $\mathbf f$ is the right-hand side of the HK8 model \eqref{eq:HK8}, and we choose 
\begin{equation} \label{eq:Phi}
\Phi = 1 + \tfrac{1}{\R} \left( 2\theta_{02}+ 4 \theta_{04} \right),
\end{equation}
so that $\ov{\Phi} = N$ according to \eqref{eq:Nu3}. As discussed in \S\ref{sec:ROM}, the two expressions \eqref{eq:Nu2} and \eqref{eq:Nu3} for the Nusselt number are equivalent along all time-averaged trajectories. Upper bounds on the two quantities proved using the auxiliary function method are also identical. To see this, notice that if $V_0 := \theta_{02}/2 + \theta_{04}/4$, then 
\begin{equation}
\tfrac{k}{4} \left( \psi_{11} \theta_{11} - \psi_{12} \theta_{12} \right) = 2\theta_{02} + 4 \theta_{04} + \mathbf{f} \cdot \nabla V_0,
\end{equation}
where the quantity on the left-hand side is the function whose time average corresponds to \eqref{eq:Nu2}. Therefore, if an upper bound on \eqref{eq:Nu2} is obtained with the auxiliary function $V$, the same bound on \eqref{eq:Nu3} can be established with the auxiliary function $V + V_0$. Solving \eqref{eq:Opt} with $\Phi$ defined by~\eqref{eq:Phi} and an auxiliary function $V$ of polynomial degree $d$ provides an upper bound ${N^*\le U^*_d}$. We performed such computations with $d=2$, 4, 6, and 8, for various values of the model parameters ($\R$, $\sigma$, and $k$). In \S\ref{sec:maxK}, we maximize these upper bounds over $k$ to provide an analogy to the maximal heat transport of the PDE. 

When solving the optimization problem in~\eqref{eq:SDP} to find the bound $U^*_d$, we do not need to consider a fully general polynomial ansatz for $V$ because some structure of $V$ can be anticipated by examining the structure of the HK8 model. Restricting the $V$ ansatz accordingly improves numerical conditioning and reduces computational cost. One source of structure is symmetry. The ODE \eqref{eq:HK8} and the quantity to be bounded~\eqref{eq:Phi} each are invariant under both of the following sign symmetries: 
\begin{equation}
\label{eq:Symm}
\begin{aligned}
&(\psi_{11}, \psi_{01}, \psi_{12}, \theta_{11}, \theta_{02}, \theta_{12}, \psi_{03}, \theta_{04}) \mapsto (\psi_{11}, -\psi_{01}, -\psi_{12}, \theta_{11}, \theta_{02}, -\theta_{12}, -\psi_{03}, \theta_{04}),\\
&(\psi_{11}, \psi_{01}, \psi_{12}, \theta_{11}, \theta_{02}, \theta_{12}, \psi_{03}, \theta_{04}) \mapsto (-\psi_{11}, \psi_{01}, -\psi_{12}, -\theta_{11}, \theta_{02}, -\theta_{12}, \psi_{03}, \theta_{04}).
\end{aligned}
\end{equation}
We impose these same symmetries on the $V$ ansatz since this does not change the optimal bounds $U^*_d$~\cite{Goluskin2019, Lakshmi2020}. The second structural constraint on $V$ comes from the requirement that the highest-degree terms in the polynomial ${\mathbf{f} \cdot \nabla V}$ be of even degree---a necessary condition for the SOS constraint in~\eqref{eq:SDP} to be satisfied. In general one expects an odd maximum degree of ${d+1}$ since $\mathbf{f}$ is quadratic. To avoid this we require that the highest-degree terms cancel in ${\mathbf{f} \cdot \nabla V}$. This imposes linear constraints on the highest-degree terms in $V$ that we encode into the $V$ ansatz. Restricting $V$ with these linear constraints and with the symmetries \eqref{eq:Symm}, we formulated the SOS problems of the form~\eqref{eq:Opt} as SDPs with the MATLAB software YALMIP~\cite{Lofberg2004, Lofberg2009} (version R20190425). The resulting optimization problems were then solved using MOSEK version 9.0.98~\cite{mosek}. Further details of our computational implementation are given in~\ref{sec:NumProc}.

\begin{figure}[tp]
\centering
\begin{tikzpicture}
\node at (0,0) {\includegraphics[scale=.5]{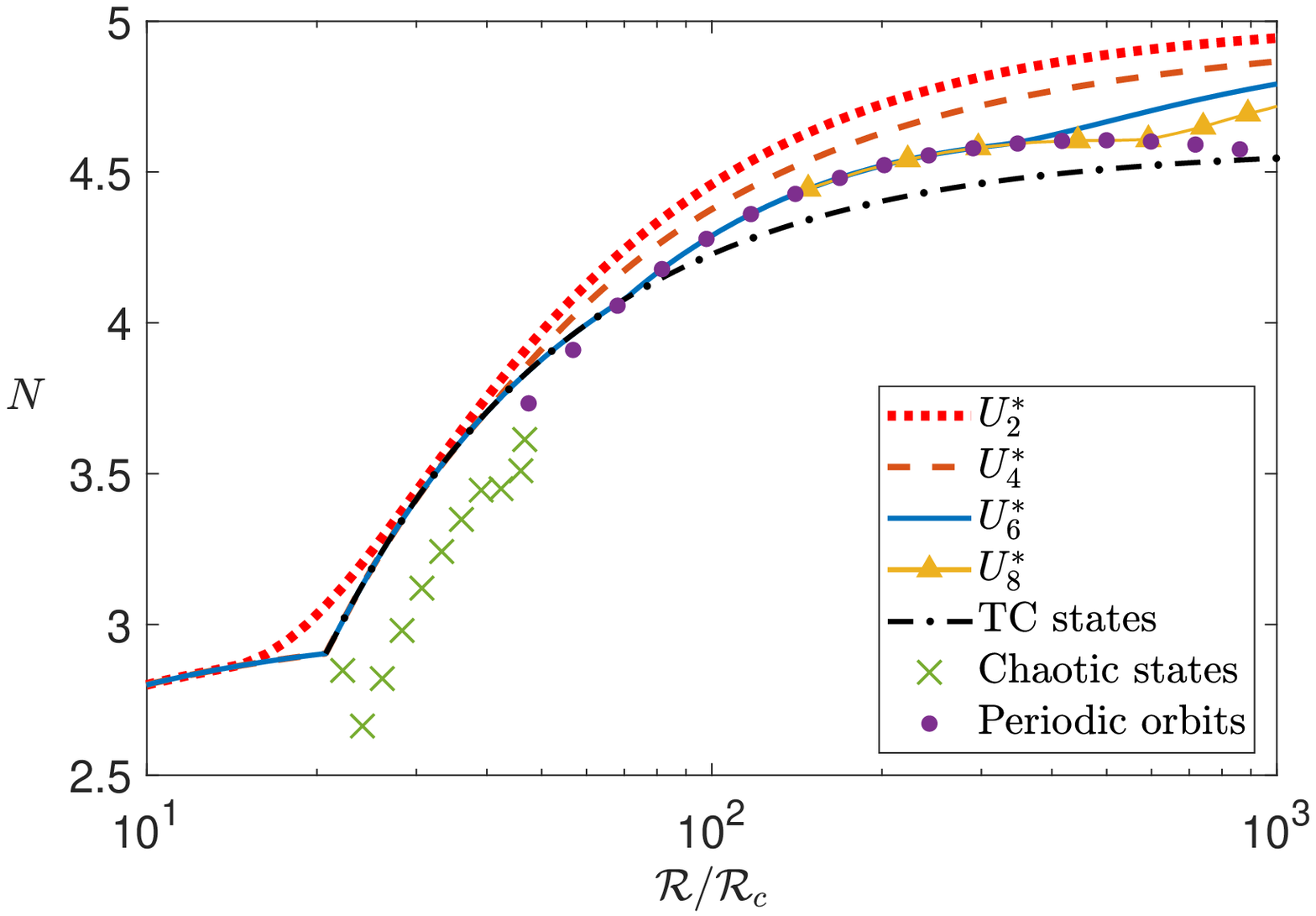}};
\node at (-5.6,2.61) {\small (a)};
\end{tikzpicture}
\begin{tikzpicture}
\node at (0,0) {\includegraphics[scale=.5]{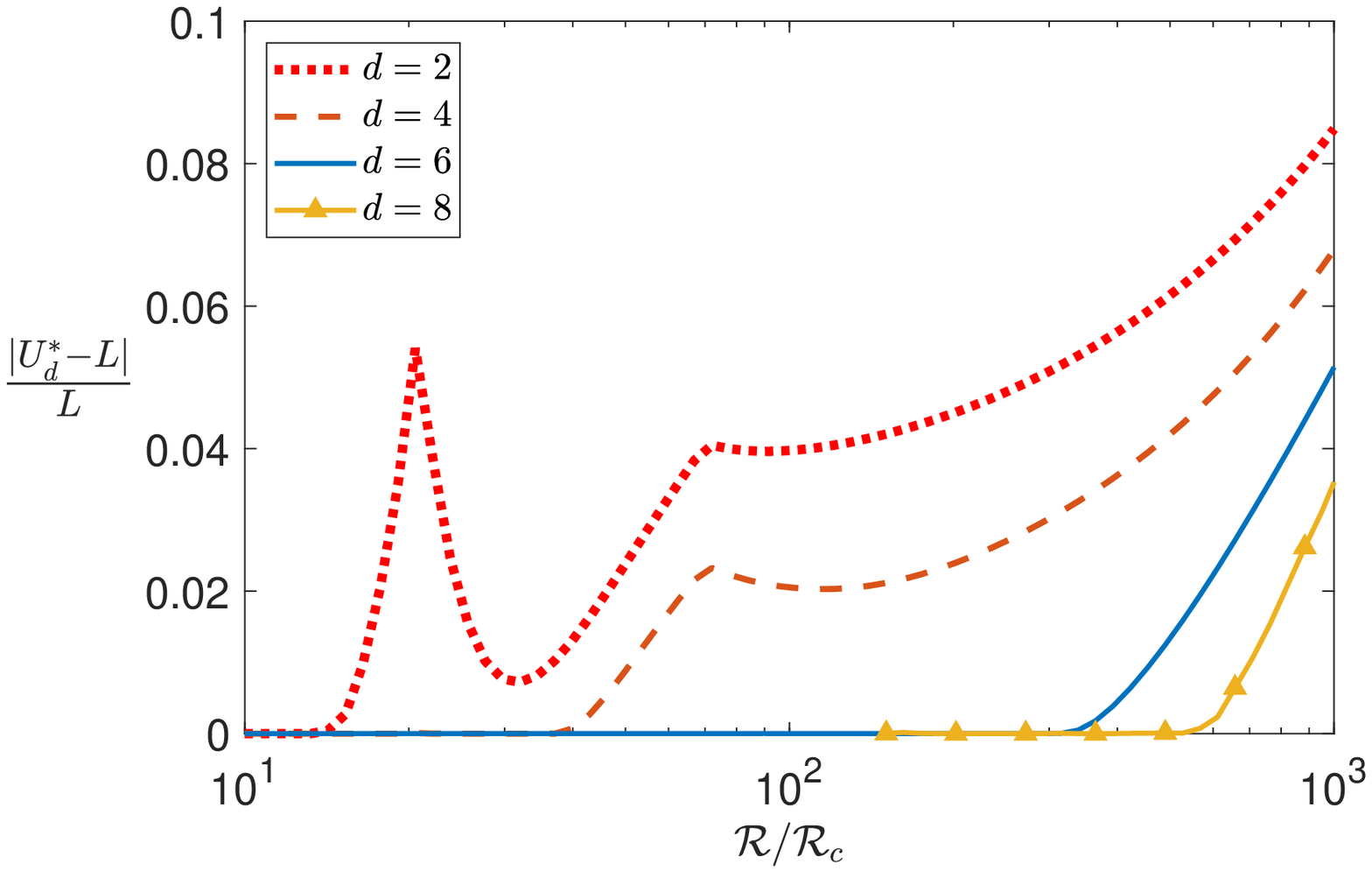}};
\node at (-5.35,2.52) {\small (b)};
\end{tikzpicture}
\caption{(a) Upper bounds ($U^*_d$) on the truncated Nusselt number ($N$) computed by solving the SOS optimization~\eqref{eq:SDP} with degree-$d$ auxiliary polynomials, compared to $N$ on particular solutions of the HK8 model with $k^2 = 1/2$ and $\sigma = 10$. (b) Relative difference between the $U^*_d$ and the lower bound $L$, determined by finding the maximum $N$ over the particular solutions obtained in \S\ref{sec:particular}.} 
\label{fig:stdCmp}
\end{figure}

As a first example we fix $(k^2 = 1/2, \sigma = 10)$, and consider the dependence of the upper bound on $\R$. At this value of $k$, the Rayleigh number $\R_{L_1}(k)$  that emerges as the first instability of the zero state takes its minimal value of $\R_c = 27/4$. Figure~\ref{fig:stdCmp}(a) shows the upper bounds we computed in this case using SOS methods. Also shown are lower bounds on $N^*$ found by searching among various trajectories of the HK8 system, as discussed in \S\ref{sec:particular}. Agreement of upper and lower bounds on $N^*$ to within numerical precision implies that the upper bounds are sharp or very nearly so, and that the corresponding trajectories maximize $N$. In such cases we say that the maximal solutions saturate the upper bounds. The relative gap between the upper bounds and lower bounds on $N^*$ established in this work is depicted in Figure~\ref{fig:stdCmp}(b); for ${\R\lesssim 560 \, \R_c}$, these two quantities agree to at least five significant digits.

Different trajectories saturate the upper bounds on $N$ over various $\R$ intervals. When ${\R\le\R_{L_1}}$, all trajectories satisfy $N=1$ since they tend to the equilibrium at the origin. On the subsequent interval ${\R_{L_1}\le\R\le\R_{TC_1}\approx 20.8 \, \R_c}$, the $L_1$ equilibria maximize $N$. At the present $k^2$ and $\sigma$ values, $N_{TC} > N_{L_1}$ for all $\R > \R_{TC_1}$, and $TC$ equilibria are maximal on the interval ${\R_{TC_1}\le\R\lesssim 69 \, \R_c}$. Time-dependent states appear to maximize $N$ for larger $\R$, with periodic orbits saturating the upper bound on the interval $69 \, \R_c \lesssim \R \lesssim 520 \, \R_c$. We draw this conclusion because $N$ on the periodic orbits agrees with the best upper bound to within the numerical error of our SOS computations. In such cases we say for simplicity that the periodic orbit saturates the bound. Strictly speaking we do not expect periodic orbits to exactly saturate a bound computed with $V$ of any finite degree $d$, as explained in~\cite{Goluskin2018}, but we ignore this distinction provided $U^*_d$ is sufficiently converged to the large-$d$ limit. The branch of periodic orbits that saturates the upper bound is the one that emerges, initially unstable, from the Hopf bifurcation at ${\R_{H_1} \approx 21.80 \, \R_c}$. For $\R$ larger than $520 \, \R_c$, time-dependent solutions may still maximize $N$, but our upper bounds on $N^*$ are not sharp enough to confirm it.

\subsection{Dependence on wavenumber and Prandtl number}
\label{sec:param}

\begin{figure}[tp]
\begin{center}
\includegraphics[scale=.5]{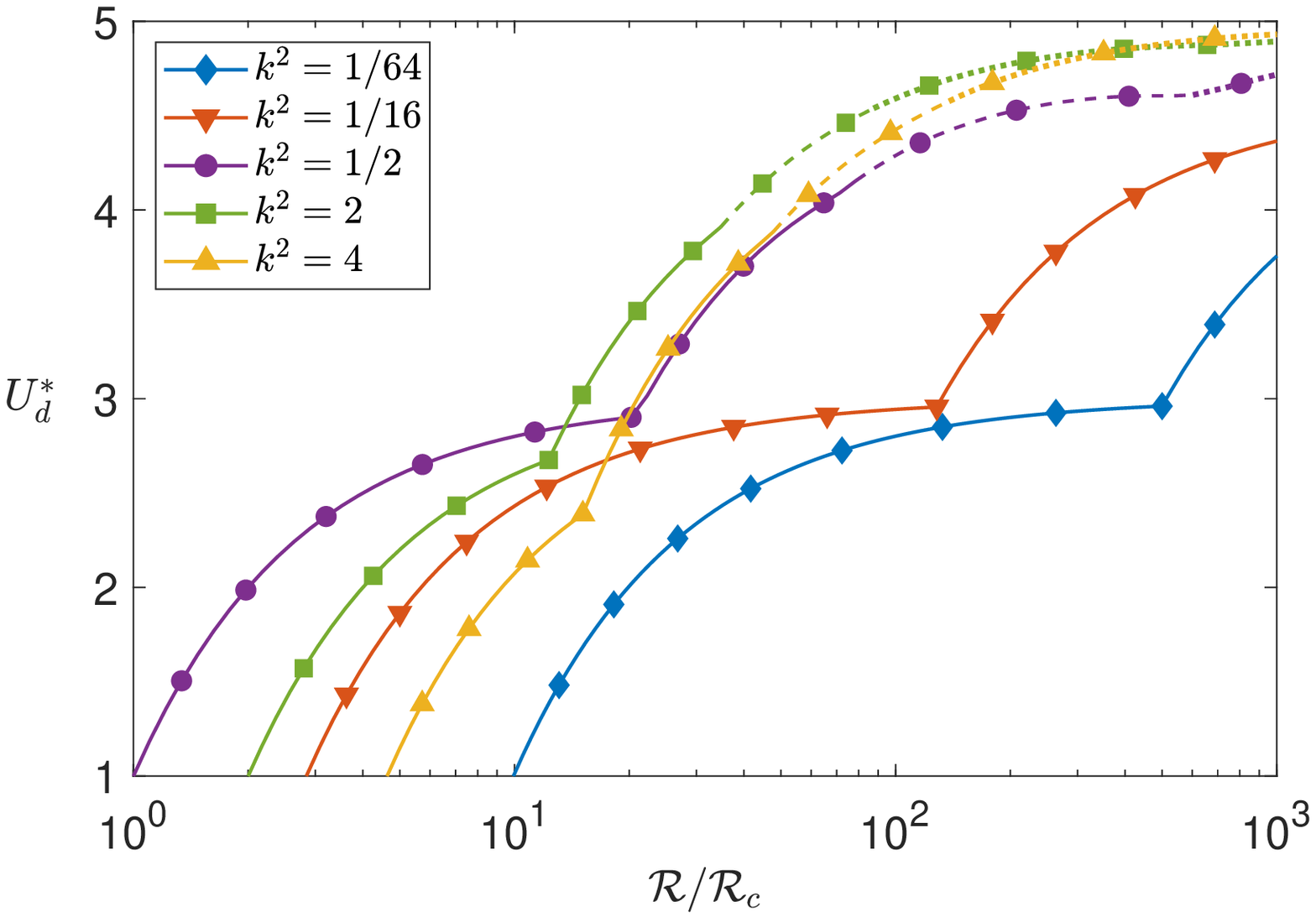}
\end{center}
\caption{Upper bounds on $N$ computed by solving the polynomial optimization problem \eqref{eq:SDP} with $V$ of degree up to eight. Upper bounds were computed for various $k$ with $\sigma=10$ across a range of $\R$. The line style indicates whether the upper bound is saturated by $L_1$ or $TC$ equilibria (\solidrule) or a time-periodic orbit (\dashedrule). The dotted lines (\dottedrule) indicate where the upper bounds of degree eight do not match the maximal $N$ among the known steady and time-dependent solutions.}
\label{fig:SDP_k}
\end{figure}

We now examine how the upper bounds and the states that saturate them depend on the parameters $k^2$ and $\sigma$. When $k^2$ is fixed to values other than $1/2$ with $\sigma$ still fixed at 10, the bounds are qualitatively similar to those depicted in Figure~\ref{fig:stdCmp}. We computed upper bounds on $N^*$ at various wavenumbers and searched among known trajectories for the largest $N$ values. Figure~\ref{fig:SDP_k} shows the upper bounds we computed at five different wavenumbers using $V$ of degrees up to eight. The different line styles in Figure~\ref{fig:SDP_k} indicate the type of state that appears to saturate the upper bounds at various $\R$ and $k$. As in the ${k^2=1/2}$ case, each bound is saturated by $L_1$ equilibria at small $\R$, by $TC$ equilibria at larger $\R$, and---at least in the smaller-$k$ cases---by periodic orbits at still larger $\R$. The fact that $L_1$ states maximize $N$ at onset is proved analytically in~\cite{Souza2015} and in~\S\ref{sec:Quad} below. The Rayleigh number where the $TC$ branch of equilibria bifurcates from the $L_1$ branch changes with $k^2$ according to \eqref{eq:Rs}, but in each case the emerging $TC$ states saturate the bound for some interval of Rayleigh number. 

Similarly, we may consider how the situation changes when $\sigma$ is fixed to various values while $k^2 = 1/2$. The analytical bound proved in~\S\ref{sec:QuadBounds1} below implies that for $k^2 = 1/2$, the $\sigma$-independent $L_1$ states maximize $N$ for $\R$ at least as large as $\R'(2^{-1/2}) = 14 \, \R_c$, where $\R'(k)$ is defined by \eqref{eq:Rprime} below. For $\R > \R'$, we computed numerical upper bounds on $N$ for various ${\sigma \in [0.01, 100]}$. In each case, the upper bounds appear to be saturated by time-dependent solutions for all $\R$ sufficiently large. The main qualitative distinction between different $\sigma$ values is whether or not there exists an interval of $\R$, starting at $\R'$, over which $TC$ states saturate the upper bounds. It was observed in~\S\ref{sec:particular} that $N_{TC}$ exceeds $N_{L_1}$ if and only if $(k^2,\sigma)$ lies in Region V of Figure~\ref{fig:bif}, which corresponds to ${\sigma \gtrsim 3.523}$ when $k^2=1/2$. Indeed, for various $\sigma > 3.523$ our upper bounds are saturated by $TC$ states over bounded intervals of $\R$, whereas for smaller $\sigma$ we found no such intervals.

\subsection{Optimal wavenumbers}
\label{sec:maxK}

In the full PDE model~\eqref{eq:BE} of 2D Rayleigh--B\'enard convection, steady solutions of each horizontal period exist for sufficiently large $\R$. Thus for the PDE it is natural to search among all horizontal periods for the steady states that maximize heat transport. The analogue in the HK8 model is to maximize $N$ over $k$---the horizontal wavenumber of modes that are included in the truncated Galerkin expansion~\eqref{eq:modes}. Thus we consider the quantity
\begin{equation} 
\label{eq:maxKOpt}
N^*_{k^*}(\R,\sigma) := \sup_{k>0} N^*(\R,k,\sigma).
\end{equation}
To find upper bounds on $N^*_{k^*}$, we repeatedly solved the SOS optimization in~\eqref{eq:SDP} using $V$ of degree six, sweeping through $k$ and using the MATLAB function \texttt{fminsearch} to converge to the global maximum of $U^*_6$ over $k$.
Lower bounds on $N^*_{k^*}$ were computed by maximizing $N$ over $k$ among each of the three types of states seen to maximize $N$ at various parameters: the $L_1$ equilibria, the $TC$ equilibria, and stable periodic orbits found by time integration. For the $L_1$ states, the optimal wavenumber is $(k^*)^2 = 1/2$, corresponding to steady convection cells of aspect ratio $2 \sqrt{2}$. Here $N_{L_1}$ attains its maximum of
\begin{equation}
\label{eq:maxL1}
\max_{k>0}N_{L_1}= 3 - \frac{27}{2 \R},
\end{equation}
in the regime $\R > \R_c$ where these $L_1$ states exist. The maximum of $N_{TC}$ over $k$ was found using Mathematica by numerically optimizing the analytical expressions for the $TC$ equilibria at various fixed values of $\R$ and $\sigma$. At these same $\R$ and $\sigma$, the maximum value of $N$ over $k$ for the stable periodic orbits was determined using a search algorithm similar to the one used to maximize the upper bound, with numerical integration performed to determine $N$ at each iteration.

\begin{figure}[tp]
\centering
\includegraphics[scale=.5]{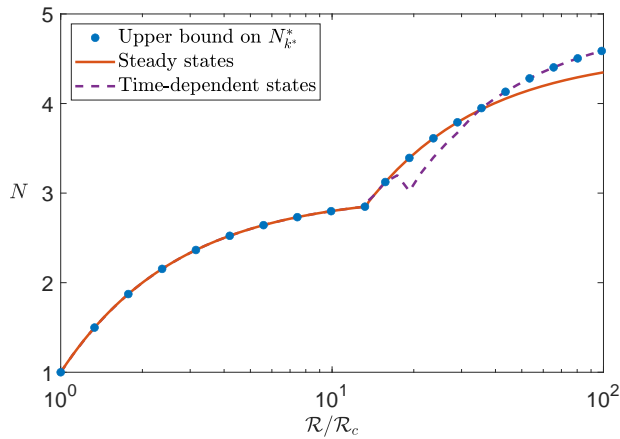}
\caption{Upper bounds on $N^*$, largest $N$ values among steady states, and largest known $N$ values among time-dependent states, each maximized over $k$ at $\sigma = 10$. The maximizer $k^*$ generally depends on $\R$ and the type of trajectory that saturates the bound. Upper bounds were computed using $V$ of degree six.}
\label{fig:maxK_sig10}
\end{figure}

Figure~\ref{fig:maxK_sig10} shows the upper and lower bounds on $N^*_{k^*}$ in the ${\sigma=10}$ case. The upper bounds are sharp or nearly sharp over the full range of $\R$. At $\sigma = 10$, the states that saturate or nearly saturate these bounds are the $L_1$ equilibria at small $\R$, the $TC$ equilibria at intermediate $\R$, and the stable periodic orbits at larger $\R$. In particular, the maximum value $N^*_{k^*}$ is attained by $N_{L_1}$ when $\R \lesssim 13.2 \, \R_c$ and steady states when ${\R \lesssim 34.2\, \R_c}$. The optimizer $k^*$ increases with $\R$, and $(k^*)^2 > 1.56$ whenever $TC$ equilibria or time-dependent states saturate the bound. Similar behavior was observed for various choices of $\sigma \in (3.5,100]$. When $\sigma < 3.5$, the value of $N_{TC}$ lies below $N_{L_1}$ (see Figure \ref{fig:bif}), and hence at small Prandtl number there is no interval of $\R$ where $TC$ saturates the upper bound on $N^*_{k^*}$.

\begin{figure}[tp]
\centering
\hspace*{-.15in}\begin{tikzpicture}[>=stealth, line width=1pt]
\node[anchor=south west, inner sep=0] at (0,0) {\includegraphics{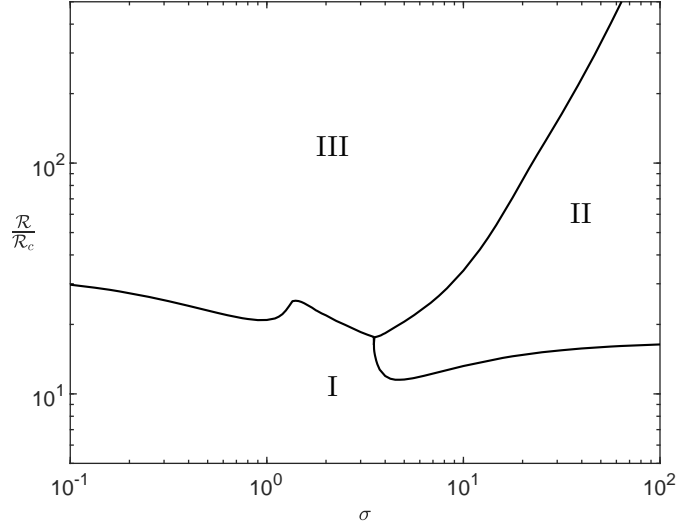}};
\node[scale=1.1] at (4.3,1.8) {I};
\node[scale=1.1] at (4.3,5) {III};
\node[scale=1.1] at (7.6,4.1) {II};
\end{tikzpicture}
\caption{Regions in the $\sigma$--$\R$ plane where the upper bound on $N^*_{k^*}$---the maximum heat transport over
$k$---is saturated by: (I) $L_1$ equilibria and (II) $TC$ equilibria. In region (III), time-periodic states appear to
saturate the upper bounds, but for some parameter combinations the upper bounds with $V$ of degree six are not sufficient to confirm this. The intersection between the three regions occurs near $\sigma = 3.5$ and $\R = 17.6 \, \R_c$,
corresponding to the minimal $\sigma$ where the Nusselt number of the $TC$ equilibria exceed that of $L_1$ for
sufficiently large $\R$.}
\label{fig:maxK_Rtrans}
\end{figure}

To study the effects of changing $\sigma$, at various fixed $\sigma$ we determined the intervals of $\R$ where our $k$-maximized upper bounds are saturated by the $L_1$ equilibria, $TC$ equilibria, or time-periodic states. Figure~\ref{fig:maxK_Rtrans} summarizes the results. At each $\sigma$, we used a bisection search to find the largest $\R$ such that an $L_1$ equilibrium saturates the $k$-maximized upper bound. That is, we computed the $k$-maximized upper bound and increased $\R$ when the upper bound differed from the maximum of $N_{L_1}$ by less than $10^{-5}$, and we decreased $\R$ otherwise. Analogous computations for $TC$ equilibria were carried out to find the boundary between regions II and III.

Regions I and II together in Figure~\ref{fig:maxK_Rtrans} comprise the parameter regime in the HK8 model where, according to our upper bounds, the maximum of $N$ over all $k$ is attained by steady states. It is an open question whether steady states maximize heat transport in the full PDE model of Rayleigh--B\'enard convection \cite{Wen2020}. For the HK8 model at all $\sigma$ and all $\R$ values small enough for the model to capture PDE behavior, Figure~\ref{fig:maxK_Rtrans} suggests that this is indeed the case.

\section{Analytical upper bounds using quadratic auxiliary functions}
\label{sec:Quad}
In principle the bounding framework~\eqref{eq:SDP} can be applied numerically or analytically, but many of the bounding computations reported in~\S\ref{sec:Numerical SDP} would be analytically intractable because the polynomial expression \eqref{eq:S} for $S$ has hundreds or thousands of terms. Bounds can be derived analytically in the case of quadratic $V$, however, and we do so in this section. The resulting bounds are rigorous and depend analytically on the parameters $\R,k,$ and $\sigma$, whereas the numerical bounds in~\S\ref{sec:Numerical SDP} were subject to rounding errors in the solutions to the SDPs, and they had to be computed anew for each triplet of parameter values.

\begin{figure}[tp]
\centering
\includegraphics[scale=1]{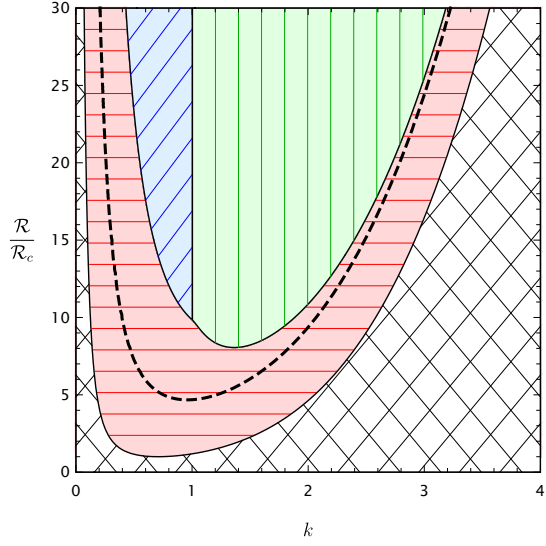}
\caption{Parameter regimes where the four different analytical upper bounds~\eqref{eq:quadBoundTotal} on $N$ are proved for the HK8 model. In the cross-hatched region, $N=1$ for all solutions. The bound $N\le N_{L_1}$ is proved here for the full horizontally hatched region; it was proved in~\cite{Souza2015} only for the part of this region below the dashed line (\dashedrule). The diagonally and vertically hatched regions correspond to the third and fourth cases in~\eqref{eq:quadBoundTotal}, respectively.}
\label{fig:quadRegion}
\end{figure}

The best analytical upper bounds on $N$ take different forms in four different regimes of the $k$--$\R$ parameter plane. These four regimes are shown in Figure~\ref{fig:quadRegion}, and the bounds in each are
\begin{equation} \label{eq:quadBoundTotal}
N \leq \begin{cases}
1, & 0 \leq \R \leq \R_{L_1}, \\[6pt]
N_{L_1}, & \R_{L_1} < \R \leq \R', \\[6pt]
N_{L_1} + \frac{1}{\R} \left[\R_{L_1} - \R_{L_2} + \sqrt{2} \sqrt{(\R-\R_{L_1})^2 + (\R - \R_{L_2})^2 } \right], \,& \R > \R', ~\, 0< k \le k', \\[6pt]
N_{L_1} + \frac{1}{\R} \left[\R_{L_1} - \R_{L_2} + \sqrt{(\R_{L_2} - \R_{L_1})^2 + 4(\R - \sqrt{\R_{L_1} \R_{L_2}})^2} \right], &  \R > \R', ~\, k > k',
\end{cases}
\end{equation}
where we recall that $\R_{L_1}(k)$ and $\R_{L_2}(k)$ are defined by \eqref{eq:RL1} and \eqref{eq:RL2} and that $N_{L_1}=3 -  2 \, \frac{\R_{L_1}}{\R}$, and where $\R'$ is defined by
\begin{equation} 
\label{eq:Rprime}
\R'(k) := \begin{cases}
\frac{1}{2} \left( \R_{L_1} + \R_{L_2} \right), & 0 \leq k \leq k',\\[6pt]
\frac{-15}{2(5k^2 - 4)} \R_{L_1} + \sqrt{\frac{11+5k^2}{5k^2-4} \R_{L_1} \R_{L_2} }, & k > k',
\end{cases}
\end{equation}
and $k' \approx 1.00319$ is the positive real root of $(5k^2+11) \R_{L_1} = (5k^2-4)\R_{L_2}$. Note that the bounds~\eqref{eq:quadBoundTotal} are uniform in $\sigma$, unlike the bounds reported in \S\ref{sec:Numerical SDP} that were computed numerically with $V$ of degree 4 and higher.

The bound in the first regime of~\eqref{eq:quadBoundTotal} is sharp and is saturated by the zero equilibrium, which is globally attracting below the first instability at $\R_{L_1}$. The bound in the second regime of~\eqref{eq:quadBoundTotal} is saturated by the $L_1$ equilibria. The same bound was proved by Souza \& Doering \cite{Souza2015} on the strictly smaller parameter regime where $\R_{L_1} <\R \le \sqrt{\R_{L_1} \R_{L_2}}$. In \S\ref{sec:QuadBounds1} we strengthen their result by extending its applicability up to the larger parameter value $\R'$. The bound in the third regime of~\eqref{eq:quadBoundTotal} is new and is proved in \S\ref{sec:QuadBounds1}. The bound in the fourth regime was proved by Souza \& Doering for ${\R > \sqrt{\R_{L_1} \R_{L_2}}}$. Their results are superseded in the second and third regimes by our new bounds. It is possible to improve the bound in the fourth regime using quadratic $V$, as suggested by the bounds we have computed numerically, but we were unable to derive an analytical expression; partial analytical results are given in~\ref{sec:QuadBounds3}. Bounds in the third and fourth regimes of~\eqref{eq:quadBoundTotal} are not sharp; it is evident from the numerical bounds reported in the previous section that $V$ of higher polynomial degrees provide better bounds.

\subsection{Sum-of-squares construction in the quadratic case}

The quadratic ansatz that we consider for the auxiliary function $V$ need not be the most general possible since some structure can be anticipated, as described in~\S\ref{sec:Numerical SDP}. First, we require that $V$ is invariant under the two symmetries~\eqref{eq:Symm} because this does not change the optimal bound. Second, we require that the cubic terms of $\mathbf{f}\cdot\nabla V$ cancel so that the polynomial $S$ contains only terms of degree two and lower. The most general quadratic $V$ satisfying these two conditions takes the form
\begin{equation}
\label{eq:Vdef}
V = c_1 \theta_{02} + c_2 \theta_{04} + c_3 \psi_{11}^2 + c_4 \psi_{01}^2 + c_5 \psi_{12}^2 + c_6 \theta_{11}^2 + c_7 \theta_{02}^2 + c_8 \theta_{12}^2 + c_9 \psi_{01} \psi_{03} + c_{10} \psi_{03}^2 + c_{11} \theta_{04}^2,
\end{equation}
where the coefficients must satisfy the linear relations:
\begin{equation}
\label{eq:quadCons}
\begin{gathered}
c_6 = c_8, \qquad c_7 = 2 c_6, \qquad 6 c_4 - c_9 - \frac{4(3+k^2)}{k^2+1} c_3 + \frac{4 k^2}{k^2+4} c_5 = 0,\\[6pt]
 \qquad c_{11} = 2c_8, \qquad \frac{2}{3} c_{10} - c_9 - \frac{4(k^2-5)}{k^2+1} c_3 + \frac{4(k^2-8)}{k^2+4} c_5 = 0.
\end{gathered}
\end{equation}
The previous derivation of upper bounds on $N$ for the HK8 model in \cite{Souza2015} was presented as an analogue of the ``background method" for PDEs~\cite{Doering96, Whitehead2011}. In the PDE setting, the background method can be viewed as a special case of a more general auxiliary functional method---the PDE analogue of our general approach~\eqref{eq:SDP}---where the auxiliary functional is quadratic~\cite{Chernyshenko2017, Goluskin2019}. Likewise, the argument in~\cite{Souza2015} is equivalent to a special case of our present analysis where the quadratic ansatz~\eqref{eq:Vdef} for $V$ has only two free coefficients (the ``background values") rather than the six free coefficients in \mbox{\eqref{eq:Vdef}--\eqref{eq:quadCons}}. \ref{sec:SD} gives the exact constraints on these coefficients that, if added, would make our analysis equivalent to~\cite{Souza2015}. We do not impose these unnecessary constraints here, leading to better bounds on $N$ in some parameter regimes.

With the quadratic $V$ ansatz~\eqref{eq:Vdef} and coefficients constrained by~\eqref{eq:quadCons}, the expression~\eqref{eq:S} for the polynomial $S$ that must be SOS becomes
\begin{equation}
\label{eq:Squad}
\begin{aligned}
 S =\;\,& U - 1 + \left(4c_1 - \frac{2}{\R}\right) \theta_{02}  + 16 c_8 \theta_{02}^2 + \left(16c_2-\frac{4}{\R}\right) \theta_{04} + 64 c_8 \theta_{04}^2\\[6pt]
 & + 2 \sigma (k^2+1) c_3 \psi_{11}^2 + k \left(-\frac{1}{2} c_1 - 2 \R c_8 - 2 \frac{\sigma}{k^2+1} c_3 \right) \psi_{11} \theta_{11} +  2(k^2+1) c_8 \theta_{11}^2 \\[6pt]
 & +  2 \sigma (k^2+4) c_5 \psi_{12}^2 + k\left(c_2 + 2\R c_8 + 2 \frac{\sigma}{k^2+4} c_5 \right) \psi_{12} \theta_{12} + 2 (k^2+4) c_8 \theta_{12}^2 \\[6pt]
 & + \frac{\sigma}{3} \left(c_9 + 4 \frac{3+k^2}{k^2+1} c_3 - 4 \frac{k^2}{k^2+4} c_5 \right) \psi_{01}^2 + 10 \sigma c_9 \psi_{01} \psi_{03} \\[6pt]
 & + 27 \sigma \left(c_9 + 4 \frac{k^2-5}{k^2+1} c_3 - 4 \frac{k^2-8}{k^2+4} c_5 \right) \psi_{03}^2.
\end{aligned}
\end{equation}
For each $(\R,k,\sigma)$, the SOS optimization~\eqref{eq:Opt} asks for the smallest $U$ such that the $c_i$ can be chosen to make the above expression an SOS polynomial of the state variables. The corresponding value of $U$ provides an upper bound on the time-averaged Nusselt number $N$ over all solutions to the HK8 model. To proceed analytically, we consider the SDP~\eqref{eq:SDP} that is equivalent to the SOS optimization~\eqref{eq:Opt}. In this formulation, the SOS constraint on expression~\eqref{eq:Squad} for $S$ is replaced by the equivalent constraint that $S=\mathbf{b}^\mathsf{T}\Q\mathbf{b}$ for some positive semidefinite Gram matrix $\Q$ and vector $\mathbf{b}$ of polynomial basis functions.

We first choose a vector $\mathbf{b}$ such that $S=\mathbf{b}^\mathsf{T}\Q\mathbf{b}$ holds for at least one matrix $\Q$, then we determine when $\Q$ can be positive semidefinite. The Gram matrix representation of $S$ is possible if and only if $S$ lies in the span of the scalar polynomial entries of the matrix $\mathbf{b}\mathbf{b}^\mathsf{T}$. Any such $\mathbf{b}$ suffices; the existence of a positive semidefinite $\Q$ does not depend on the choice of $\bb$. Here we simply choose the entries of $\bb$ to be monomials:
\begin{equation}
\bb=\begin{bmatrix}\bb_1\\\bb_2\\\bb_3\\\bb_4\end{bmatrix}, 
\quad \text{where} \quad
\mathbf{b}_1 =
 \begin{bmatrix} \psi_{11} \\ \theta_{11} \end{bmatrix}, \quad \mathbf{b}_2  = 
 \begin{bmatrix} \psi_{01} \\ \psi_{03} \end{bmatrix}, \quad
\mathbf{b}_3 = 
 \begin{bmatrix} \psi_{12} \\ \theta_{12} \end{bmatrix},
\quad \mathbf{b}_4 =
 \begin{bmatrix} 1 \\ \theta_{02} \\ \theta_{04} \end{bmatrix}.
\end{equation}
We have grouped the entries of $\bb$ into the four sub-vectors $\bb_i$ to exploit symmetry. In particular, because the expression~\eqref{eq:Squad} for $S$ is invariant under both transformations in~\eqref{eq:Symm}, we group monomials such that $\bb_1$, $\bb_2$, $\bb_3$, and $\bb_4$ are invariant under, respectively, the first transformation only, the second transformation only, neither, and both. We then restrict $\Q$ to be block diagonal with blocks $\Q_i$ sized according to the $\bb_i$. In this case the relation $S=\bb^\mathsf{T}\Q\bb$ becomes
\begin{equation}
\label{eq:blocks}
S = \sum_{i=1}^4\mathbf{b}_i^\mathsf{T} \Q_i \mathbf{b}_i,
\end{equation}
and this implies
\def\x{\begin{bmatrix}
\frac{\sigma}{3} \left(c_9 + 4 \frac{3+k^2}{k^2+1} c_3 - 4 \frac{k^2}{k^2+4} c_5 \right) & 5 \sigma c_9 \\
5 \sigma c_9 & 27 \sigma \left(c_9 + 4 \frac{k^2 - 5}{1 + k^2} c_3 - 4 \frac{k^2-8}{k^2+4} c_5 \right)
\end{bmatrix}}
\begin{equation} \label{eq:fullQ}
\begin{aligned}
\Q_1 &=\begin{bmatrix}
2 \sigma (k^2+1) c_3  & \hspace{1.4ex} -\frac{k}{2} \left(\frac{1}{2} c_1 + 2 \R c_8 + 2 \frac{\sigma}{k^2+1} c_3 \right) \\
-\frac{k}{2} \left(\frac{1}{2} c_1 + 2 \R c_8 + 2 \frac{\sigma}{k^2+1} c_3 \right) & \hspace{1.4ex} 2(k^2+1) c_8 
\end{bmatrix}, \\[6pt]
\Q_2 &= \begin{bmatrix}
\frac{\sigma}{3} \left(c_9 + 4 \frac{3+k^2}{k^2+1} c_3 - 4 \frac{k^2}{k^2+4} c_5 \right) & 5 \sigma c_9 \\
5 \sigma c_9 & 27 \sigma \left(c_9 + 4 \frac{k^2 - 5}{1 + k^2} c_3 - 4 \frac{k^2-8}{k^2+4} c_5 \right)
\end{bmatrix}, \\[6pt]
\Q_3 &= \begin{bmatrix}
2 \sigma (4 + k^2) c_5 &  \hspace{8ex} \frac{k}{2} \left(c_2 + 2 \R c_8 + 2 \frac{\sigma}{k^2+4} c_5 \right)\\
 \frac{k}{2} \left(c_2 + 2 \R c_8 + 2 \frac{\sigma}{k^2+4} c_5 \right) & \hspace{8ex} 2 (k^2+4) c_8
\end{bmatrix}, \\[6pt]
\Q_4 &= \begin{bmatrix}
\hspace{4.37ex} U - 1 &  \hspace{8.73ex} 2 c_1 - 1/\R  &  \hspace{8.73ex} 8 c_2 - 2/\R\phantom{\hspace{4.37ex}}\\
\hspace{4.37ex} 2 c_1 - 1/\R  & \hspace{8.73ex} 16 c_8  & \hspace{8.73ex} 0 \phantom{\hspace{4.37ex}}\\
\hspace{4.37ex} 8 c_2 - 2/\R  & \hspace{8.73ex} 0  & \hspace{8.73ex} 64 c_8 \phantom{\hspace{4.37ex}}
\end{bmatrix}.
\end{aligned}
\end{equation}
There is no loss of generality in letting all entries of $\Q$ outside the $\Q_i$ blocks be zero because if there exists any $\Q\succeq0$ satisfying $S=\bb^\mathsf{T}\Q\bb$, then there exists such a $\Q$ that is block diagonal~\cite{Gatermann2004}. This simplifies matters because the condition $\Q\succeq0$ is equivalent to $\Q_i\succeq0$ holding for each block. In other words, $S$ is an SOS polynomial if and only if each $\bb_i^\mathsf{T}\Q_i\bb_i$ is an SOS polynomial. To prove an upper bound $N\le U$ in the following analysis, it suffices to find coefficients $c_i$ such that $\Q_i\succeq0$ for all four matrices in~\eqref{eq:blocks}--\eqref{eq:fullQ}. A similar analytical procedure was implemented in~\cite{Goluskin2018} to exploit symmetry when bounding time averages in the Lorenz equations.

\subsection{Analytical bounds near the onset of convection} \label{sec:QuadBounds1}

The origin ceases to be globally attracting when the $L_1$ equilibria emerge as $\R$ increases past $\R_{L_1}$, corresponding to the onset of convection in the PDE model. At each $k$ there exists an interval of Rayleigh number where the $L_1$ states maximize $N$. In this subsection we prove that
\begin{equation} 
\label{eq:quadBound1}
 N \leq N_{L_1} = 1 + 2\left(1-\frac{\R_{L_1}}{\R}\right)
\end{equation}
when $\R_{L_1} \leq \R \leq \R'$, where $\R'(k)$ is defined as in~\eqref{eq:quadBoundTotal}. The regime of the $k$--$\R$ plane where this bound is proven is represented by horizontal hatching in Figure~\ref{fig:quadRegion}. The bound \eqref{eq:quadBound1} may hold at some $\R$ values larger than $\R'$ for certain $k^2$ and $\sigma$, but this cannot be proved using $V$ that are quadratic.

To prove~\eqref{eq:quadBound1}, we let $U=N_{L_1}$ in the expression~\eqref{eq:Squad} for $S$. When this bound holds it is saturated by the $L_1$ equilibria. The auxiliary function method can give a sharp bound on time averages only if $S$ vanishes pointwise on all trajectories that saturate the bound~\cite{Tobasco2018}, so in the present case $S$ must vanish on the $L_1$ equilibria, whose nonzero coordinates are given by~\eqref{eq:L1}. This is possible only if all four terms in the sum $S=\bb_i^\mathsf{T}\Q_i\bb_i$ are SOS polynomials that vanish at the $L_1$ equilibria. The second and third terms vanish there for any $\Q_i$ because $\bb_2$ and $\bb_3$ vanish. The first and fourth terms, on the other hand, vanish at the $L_1$ equilibria if and only if they take the form
\begin{align}
\bb_1^\mathsf{T}\Q_1\bb_1 &= q_1 \left[\psi_{11} - \tfrac{k}{(k^2+1)^2} \theta_{11}\right]^2, &
\bb_4^\mathsf{T}\Q_4\bb_4 &= q_{4} \left[\theta_{02} - (\mathcal{R} - \mathcal{R}_{L_1} )\right]^2 + q_5 \theta_{04}^2,
\end{align}
where the SOS constraints require $q_1, q_4, q_5 \geq 0$. Applying the above identities on the right-hand side of $S=\bb_i^\mathsf{T}\Q_i\bb_i$ and equating coefficients on each side of this equality determines four of the coefficients of $V$:
\begin{align} \label{eq:coeffs1}
c_1 &= -\frac{1}{2 \R}, &
c_2 &= \frac{1}{4 \R}, &
c_3 &= \frac{(k^2+1)^4}{8 \sigma k^2 \R(\mathcal{R} - \mathcal{R}_{L_1})}, &
c_8 &= \frac{1}{8 \R (\mathcal{R} - \mathcal{R}_{L_1})}.
\end{align}
To establish the bound \eqref{eq:quadBound1}, the eleven coefficients of the $V$ ansatz~\eqref{eq:Vdef} must satisfy not only the four expressions above but also the five constraints in~\eqref{eq:quadCons}. This ensures $\Q_1,\Q_4\succeq0$ when $\R\ge\R_{L_1}$, so it remains only to choose coefficients $c_5$ and $c_9$ such that $\Q_ 2, \Q_3 \succeq 0$. 
%These matrix inequalities can be satisfied if and only if $\R_{L_1} \leq \R \leq \R'$.
We observe that a 2-by-2 matrix is positive semidefinite if and only if its upper left entry and determinant are both nonnegative. Applying this criterion gives four inequalities that are equivalent to $\Q_2$ and $\Q_3$ being positive semidefinite. We performed quantifier elimination using the \texttt{Reduce} and \texttt{Exists} commands in Mathematica to determine that these inequalities can be satisfied if and only if $\R_{L_1} \leq \R \leq \R'$. Thus, quadratic $V$ yield the sharp bound $N^* = N_{L_1}$ on this parameter regime.

\subsection{Analytical bounds at larger Rayleigh number}
\label{sec:QuadBounds2}

In the regime where $\R \ge \R'$, we have proved a new analytical bound when $k \leq k'\approx1.00319$ but not when $k>k'$. Our bound in the former case is
\begin{equation} \label{eq:quadBound2}
N \leq 3 - \frac{2 \R_{L_1}}{\R} + \frac{1}{\R}\left[ \R_{L_1} - \R_{L_2} + \sqrt{2}\sqrt{(\R- \R_{L_1})^2 + (\R- \R_{L_2})^2} \right].
\end{equation}
To derive this bound we consider the expression \eqref{eq:Squad} for $S$ where $U$ is equal to the right-hand side of~\eqref{eq:quadBound2}. As in the previous subsection, we must show that the resulting $S$ can be written in the form $S=\sum_{i=1}^4\mathbf{b}_i^\mathsf{T} \Q_i \mathbf{b}_i$, where each term in the sum is an SOS polynomial---or, equivalently, where each $\Q_i\succeq0$.

Allowing for fully general SOS constraints leads to analytical difficulties, even with the simplifying block diagonal structure of $\Q$. Unlike in \S\ref{sec:QuadBounds1}, we cannot anticipate where the polynomial $S$ must vanish. Instead we simplify the analysis by making assumptions on the forms of the SOS representations. In particular we observe that, in the $\sigma \to \infty$ limit, the Lorenz triplets $\{\psi_{11},\theta_{11},\theta_{02}\}$ and $\{\psi_{12},\theta_{12},\theta_{04}\}$ decouple, and the maximal $N$ is obtained when each triplet is in the $L_1$ and $L_2$ states, respectively (see \ref{sec:InfPr} for details). This motivates us to assume that the SOS representations of the first and third SOS polynomials take the form
\begin{align} 
\label{eq:basis2}
\mathbf{b}_1^\mathsf{T} \Q_1 \mathbf{b}_1 &= q_1\left[\psi_{11} - \frac{k}{(k^2+1)^2} \theta_{11} \right]^2, &
\mathbf{b}_3^\mathsf{T} \Q_3 \mathbf{b}_3 &=  q_3\left[\psi_{12} + \frac{k}{(k^2+4)^2} \theta_{12} \right]^2.
\end{align}
The above constraints are stronger than the general SOS conditions and could potentially lead to suboptimal bounds, but this appears to not occur; at various fixed parameter values in this regime, upper bounds computed numerically with the optimal choice of quadratic $V$ agree precisely with the analytical bound~\eqref{eq:quadBound2}.

\begin{figure}[tp]
\centering
\begin{tikzpicture}
\node at (0,0) {\includegraphics[scale=.5]{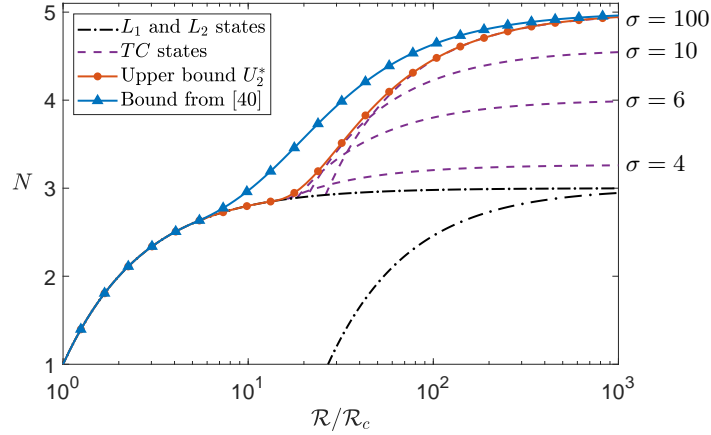}};
\node[anchor=west] (A) at (3.9,2.65) {\small $\sigma = 100$};
\node[anchor=west] (B) at (3.9,2.22) {\small $\sigma = 10$};
\node[anchor=west] (C) at (3.9,1.56) {\small $\sigma = 6$};
\node[anchor=west] (D) at (3.9,.71) {\small $\sigma = 4$};
\end{tikzpicture}
\caption{Analytical upper bounds on $N^*$ in the $k^2 = 1/2$ case, proved with optimal quadratic auxiliary functions ($U^*_2$) and with the suboptimal choice of \cite{Souza2015}. The bounds are uniform in $\sigma$. Values of $N_{TC}$ are shown at several selected values of $\sigma$ to demonstrate near-convergence of $U^*_2$ to the envelope of steady state Nusselt numbers over $\sigma$.} 
\label{fig:quadBound_std}
\end{figure}

With the assumption~\eqref{eq:basis2} on SOS representations, the coefficients of $V$ satisfy
\begin{equation} \label{eq:coeffs2}
c_3 = \frac{(k^2+1)}{\sigma} \R_{L_1} c_8, \quad c_5 = \frac{(k^2+4)}{\sigma} \R_{L_2} c_8, \quad c_1 = 4 (\R_{L_1} - \R) c_8, \quad c_2 = 2 (\R_{L_2}- \R) c_8.
\end{equation}
As a result of \eqref{eq:basis2}, the matrices $\Q_1$ and $\Q_3$ defined in \eqref{eq:fullQ} each have determinant zero, and so $\Q_1,\Q_3\succeq0$  as long as $c_8 \geq 0$. The condition $\Q_2 \succeq 0$ requires $c_8 > 0$ because $c_8 = 0$ would imply that $\det{\Q} = -16 \sigma^2 c_9^2$, which is negative for all $c_9$. Furthermore, it can be shown that if $c_8$ is positive, $\Q_4 \succeq 0$ whenever $\det{\Q_4} \geq 0$. We also observe that $\det{\Q_4} = 0$ must hold at the minimal $U$; otherwise, there would exist smaller $U$ such that $\Q_4\succeq0$.  After the relations~\eqref{eq:coeffs2} are applied to $\Q_4$, the $\det{\Q_4} = 0$ condition becomes
\begin{equation} \label{eq:Umin}
U - 1 = \frac{\left(8 (\R_{L_1} - \R) c_8 - 1/\R \right)^2 + \left(8 (\R_{L_2} - \R) c_8 - 1/\R \right)^2}{16 c_8}.
\end{equation}
Minimizing $U$ over positive $c_8$ yields the bound \eqref{eq:quadBound2}, where the minimizer is
\begin{equation}
c_8 = \left[32 \R^2 \left((\R - \R_{L_1})^2 + (\R - \R_{L_2})^2 \right) \right]^{-1/2}.
\end{equation}
It remains to find $c_9$ such that $\Q_2 \succeq 0$. Again performing quantifier elimination with Mathematica's \texttt{Reduce} and \texttt{Exists} commands, we find that such $c_9$ exist if and only if $k \leq k'$. This condition and the $\R\ge\R'$ condition define the regime in the $\R$--$k$ plane where we proved the bound~\eqref{eq:quadBound2}. For the standard wavenumber $k^2=1/2$, Figure~\ref{fig:quadBound_std} compares the optimal analytical bounds---\eqref{eq:quadBound1} and \eqref{eq:quadBound2}---that can be proved using quadratic $V$, to the upper bound from \cite{Souza2015}, as well as to the $N$ values of various steady states.

\subsection{Quadratic bounds compared to steady states at maximal Prandtl number}

The analytical bounds~\eqref{eq:quadBoundTotal} proved using quadratic $V$ are uniform in $\sigma$, so they are also upper bounds on the maximum of $N$ over $\sigma$. That is,
\begin{equation} \label{eq:NmaxS}
N^*_{\sigma^*}(k,\R) := \max_{\sigma > 0} N^*(k,\sigma,\R) \le U^*_{2}(k,\R).
\end{equation}
When $\R_{L_1} \leq \R \leq \R'(k)$, the quadratic-$V$ upper bound is saturated by the $\sigma$-independent $L_1$ states, and $N^*_{\sigma^*} = U^*_2$. For $\R>\R'$, the uniform-in-$\sigma$ bounds cannot always be sharp at particular $\sigma$ because there are cases where $N$ is maximized by the $\sigma$-dependent $TC$ states. However, this does not rule out the possibility that the uniform-in-$\sigma$ bounds may be sharp upper bounds on $N^*_{\sigma^*}$. Investigating this possibility, we find that the analytical bounds~\eqref{eq:quadBoundTotal} are nearly equal to $N^*_{\sigma^*}$ but slightly larger in general when $\R > \R'$.

\begin{figure}[tp]
\centering
\includegraphics[scale=.5]{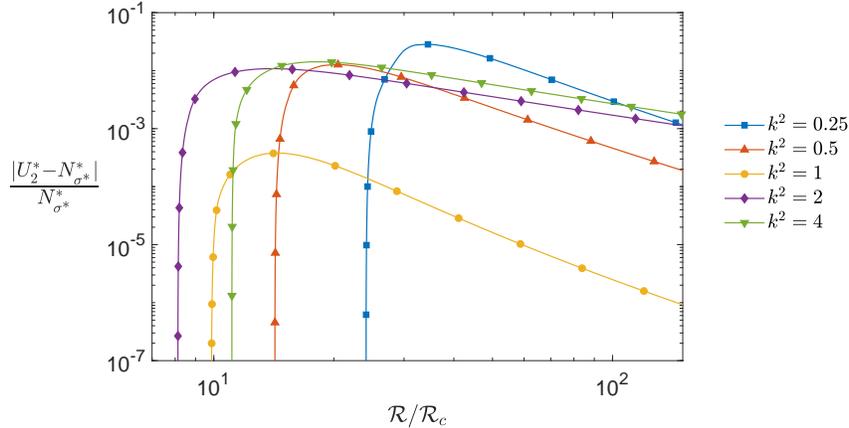}
\caption{Difference between the best upper bound provable using quadratic auxiliary functions ($U^*_2$) and the value $N^*_{\sigma^*}$ defined by~\eqref{eq:NmaxS}, for various fixed values of $k^2$. The difference vanishes as $\R$ decreases towards $R'(k)$ and in the limit $\R\to \infty$.}
\label{fig:quadBound_comp}
\end{figure}

The relationship between the quadratic-$V$ upper bounds and the quantity $N^*_{\sigma^*}$ may be visualized by constructing an envelope of $N_{TC}$ curves at multiple values of $\sigma$. Figure~\ref{fig:quadBound_std} shows a few such curves in the $k^2 = 1/2$ case. The bound $U_2^*$ follows the contour of this envelope, but a small separation occurs after the quadratic-$V$ bound diverges from $N_{L_1}$. As shown in Figure~\ref{fig:quadBound_comp} for various fixed $k$, the bounds provided by quadratic auxiliary functions are almost but not quite saturated by $N^*_{\sigma^*}$ when $\R>\R'$. The $N^*_{\sigma^*}$ values used in Figure~\ref{fig:quadBound_comp} we obtained by finding exact expressions for $N_{TC}$ with computer algebra, then maximizing the result over $\sigma$ for various fixed values of $\R$ and $k$. In each case, the maximizing $\sigma^*$ lies in region V of Figure~\ref{fig:bif}---the parameter regime in the $k^2$--$\sigma$ plane where the branch of $TC$ equilibria connects only to the $L_1$ branch. As $\R\to\infty$, the quadratic-$V$ bounds and the infinite-$\sigma$ limit of $N_{TC}$ both asymptote to $N=5$; see~\ref{sec:InfPr} for details on the infinite-$\sigma$ limit.

\section{Conclusions}
\label{sec:Con}
The auxiliary function method was applied using sum-of-squares optimization to establish upper bounds on the mean heat transport, $N$, among all solutions of the HK8 system, a truncated version of Rayleigh's PDE model~\cite{Rayleigh1916}. Values of $N$ were also calculated along various particular solutions to the HK8 system. The upper bounds are sharp in many cases, as confirmed by their coincidence with $N$ on a known steady or time-periodic solution. For purposes of numerical computation, SOS optimization was performed via semidefinite programming with auxiliary functions of polynomial degrees 2, 4, 6, and 8 at various choices of the model parameters $\R$, $\sigma$, and $k$. Moreover, upper bounds were derived analytically using quadratic auxiliary functions, yielding estimates that depend explicitly on the parameters $\R$ and $k$, and improving upon a previous result.

For all values of $k$ and $\sigma$ where we computed bounds by SOS optimization, steady states maximize $N$ for a range of $\R$ past the onset of convection. Specifically, for all $k$ and $\sigma$ there exists an interval of $\R$ where the $L_1$ equilibria maximize $N$. This interval contains the $\sigma$-independent interval $\R_{L_1} \leq \R \leq \R'$, as follows from our analytical bounds. When $\sigma$ is sufficiently large, there exists a bounded interval of $\R$ within $(\R',\infty)$ where the $TC$ equilibria saturate the upper bound; for smaller $\sigma$ the $L_1$ equilibria transport optimally among all steady states of the HK8 model. Numerical results suggest that time-dependent trajectories typically maximize $N$ for sufficiently large $\R$. The emergence of time-periodic solutions that transport more heat than any steady state contrasts with the Lorenz equations where the $L_1$ equilibrium maximizes $N$ for all $\R$ beyond onset \cite{Goluskin2018,Lorenz63,Souza2015PLA}. Motivated by the physics of the full PDE model, where the aspect ratio of convection cells need not be fixed, we also maximized our bounds over $k$. Similar maximization over the horizontal period of the primary roll state has been carried out in recent studies of maximal heat transport for the full PDE~\cite{Hassanzadeh2014,Souza2020,Wen2015}. Doing so naturally divides the $\sigma$--$\R$ parameter space into regions (Figure~\ref{fig:maxK_Rtrans}) where the so-maximized bounds are saturated by each of the three types of solutions described above. For a range of $\R$ after the onset of convection, bounds are always saturated by steady states. This means that maximal heat transport is achieved by steady flows, at least for the small
range of $\R$ where the HK8 model faithfully reflects the full PDE.

The HK8 model considered in this paper is but one of many distinguished truncations of Rayleigh's 1916 model satisfying certain conservation laws obeyed by the PDE.
A systematic approach to identifying such physically distinguished models was first explored by Thiffeault \cite{Thiffeault1995,Thiffeault1996}, who provided a guide to building higher-dimensional ODE models of Rayleigh--B\'enard convection.
It remains for future work to derive a hierarchy of such models and compute upper and lower bounds on the truncated Nusselt number in successively larger models. In principle, this can be done using the same computational methods we have illustrated here for the HK8 model.
This program may ultimately reveal solutions that maximize heat transport in the full PDE model of Rayleigh--B\'enard convection.
Analysis of increasingly faithful approximations of the PDE may illuminate whether or not steady coherent convection realizes maximal heat transport.

\vspace{4ex}

\noindent \textbf{Acknowledgments} Two of the authors (MO and WWS) were partially supported by NSF award CBET-1604903, two (MO and CRD) by NSF award DMS-1813003, and one (DG) by a Van Loo postdoctoral fellowship at University of Michigan and by the NSERC Discovery Grants Program via awards RGPIN-2018-04263, RGPAS-2018-522657, and DGECR-2018-00371.
Computational resources and services were provided by Advanced Research Computing at the University of Michigan. 

\appendix

\section{Numerical procedure}
\label{sec:NumProc}
In this appendix, we detail the computational procedure used to solve the SOS optimization problems whose results are reported in~\S\ref{sec:Numerical SDP}. If the SDP corresponding to~\eqref{eq:Opt} is solved without simplification, computational cost and numerical ill-conditioning quickly become prohibitive as the polynomial degree is raised. Both aspects may be improved by restricting the ansatz for the auxiliary function $V$. Numerical conditioning can be improved by rescaling the phase space variables in the governing ODE system. Here we detail the specific monomial reduction and rescaling used to produce the results of~\S\ref{sec:Numerical SDP}.

The HK8 ODE and the expression for $N$ are each invariant under the symmetries~\eqref{eq:Symm}. We impose this same invariance on $V$ since doing so does not affect the optimal value of the resulting SOS problem \eqref{eq:Opt} \cite{Goluskin2019, Lakshmi2020}. The first symmetry in~\eqref{eq:Symm} dictates the degrees of $\psi_{12}, \theta_{12}, \psi_{01}$, and $\psi_{03}$ have an even sum in each monomial of the $V$ ansatz, and the second symmetry in~\eqref{eq:Symm} dictates the same for $\psi_{11}, \theta_{11}, \psi_{01}$, and $\psi_{03}$. The symmetric monomials take the form:
\begin{equation}
(\psi_{01} \psi_{03})^{d_1} (\psi_{11} \theta_{11})^{d_2} (\psi_{12} \theta_{12})^{d_3} \,\theta_{02}^{d_4} \,\theta_{04}^{d_5}\, p(\bx)^2,
\end{equation}
where $d_1,\ldots,d_5$ are nonnegative integers and $p(\bx)$ is any monomial of the HK8 variables. Since $\Phi$, $V$, and the ODE share the same symmetries, the polynomial $S$ defined by~\eqref{eq:S} does also. For a properly ordered polynomial basis vector, the Gram matrix representing $S$ can then be written in block diagonal form without changing the optimum of the SDP. This block diagonalization is automated by YALMIP, and in~\S\ref{sec:Quad} it is illustrated explicitly for the case of quadratic $V$. In SDP computations, block diagonalization significantly reduces computational cost and memory usage and improves conditioning.

The $V$ ansatz can be further restricted by observing that the SOS constraint on $S$ requires the highest-degree monomials in $S$ to be of even degree. Such monomials generally come from the $\mathbf{f} \cdot \nabla V$ term in~\eqref{eq:S}. If the highest-degree monomials in $V$ are of even degree $d$, then for the HK8 model the polynomial $\mathbf{f} \cdot \nabla V$ generally includes terms of odd degree $d+1$. Hence the SOS constraint can be satisfied only if the leading terms in $V$ are constrained such that the highest-degree terms in $\mathbf{f}\cdot \nabla V$ cancel. This condition amounts to linear constraints on the coefficients of the highest-degree terms of $V$. In the present application to the HK8 model, these linear constraints imply that the degree-$d$ terms of $V$ take the form
\begin{equation}
(\psi_{01} \psi_{03})^a q(\bx)^2,
\end{equation}
where $a$ is a nonnegative integer and $q(\bx)$ is any monomial of degree $(d - a)/2$. This condition on $V$, along with the imposed symmetry, restricts $V$ to a subspace of the vector space $\mathbb{P}_{8,d}$ whose dimension is much smaller than the full dimension, as summarized in Table~\ref{tab:Monoms}. In theory the smaller $V$ ansatz gives the same bounds, but in practice it approximates these bounds with less numerical error, as well as lower computational cost.

\begin{table}[t]
\centering
\captionof{table}{Number of monomials in the ansatz for the auxiliary function $V$ of degree $d$ before and after reducing the ansatz using the structure of the HK8 model. The number of monomials before reduction is $\binom{8+d}{d}$.} \label{tab:Monoms} 
$\begin{array}{ccc}
\mbox{$d$} & \mbox{Unreduced} & \mbox{Reduced} \\
\midrule
2 & 45 & 11 \\
4 & 495 & 88 \\ 
6 & 3003 & 488 \\
8 & 12870 & 2084\\
10 & 43758 & 7251\\
\end{array}$
\end{table}

Scaling the ODE variables has a significant impact on the numerical conditioning of the SDP computations. A common heuristic implemented when using SOS optimization to study dynamical systems is to scale the state variables so that the relevant trajectories lie within the region $[-1,1]^n$~\mbox{\cite{Goluskin2018,Henrion2014}}. The appropriate variable scalings for the HK8 system were determined empirically using a combination of time integration and SOS bounds on the time average of each state
variable. To achieve the desired scaling across a wide range of parameter values, the two Lorenz triplets $\{\psi_{11}, \theta_{11}, \theta_{02}\}$ and $\{\psi_{12}, \theta_{12}, \theta_{04}\}$ were scaled by their values at the $L_1$~\eqref{eq:L1} and $L_2$~\eqref{eq:L2} steady states, respectively. The remaining variables, $\psi_{01}$ and $\psi_{03}$, were scaled by $\sqrt{\R}$ and $\sqrt{\R}/27$, respectively, motivated by their values at the $TC$ equilibria. For many computations, all variables were scaled down further, typically by a factor of two, because doing so was empirically observed to reduce the numerical error.

After defining the $V$ ansatz and rescaling the ODE variables as described above, the optimization toolbox YALMIP~\cite{Lofberg2004} (version R20190425) was used to formulate the SOS optimization problem~\eqref{eq:Opt} as an SDP~\eqref{eq:SDP} and interface with the SDP solver. The resulting SDP was solved using MOSEK~\cite{mosek} (version 9.0.98), which implements a primal-dual interior-point algorithm. Most computations were performed on a 3.0 GHz Intel Xeon processor, with some smaller problems solved on a laptop with a 2.2 GHz Intel i5 processor.

\section{Limiting cases of the HK8 model} \label{sec:Limits}
\subsection{The HK8 system in the infinite Prandtl number limit} \label{sec:InfPr}

In this appendix, we examine the HK8 model in the limit of large Prandtl number, and determine upper bounds on $N$ using SOS optimization. Proper balancing of terms in the HK8 system suggests that as $\sigma \to \infty$, the shear modes $\psi_{01}$ and $\psi_{03}$ are $O(\sigma^{-1})$, and all other variables are $O(1)$ as $\sigma \to \infty$. Scaling the state variables of the HK8 system according to these assumptions allows the dynamics to be reduced to two 2-dimensional systems:
\begin{equation} \label{eq:infPr1}
\begin{aligned}
\dot{\theta}_{11} &= \frac{k^2+1}{\R_{L_1}} \big( \R  - \R_{L_1} - \theta_{02} \big) \theta_{11},\\
\dot{\theta}_{02} &= -4 \theta_{02} + \frac{k^2+1}{2 \R_{L_1}} \theta_{11}^2,
\end{aligned}
\end{equation}
and
\begin{equation} \label{eq:infPr2}
\begin{aligned}
\dot{\theta}_{12} &= \frac{k^2+4}{\R_{L_2}} \big( \R - \R_{L_2} - 2 \theta_{04} \big) \theta_{12},\\
\dot{\theta}_{04} &= -16 \theta_{04} + \frac{k^2+4}{\R_{L_2}} \theta_{12}^2,
\end{aligned} 
\end{equation} 
where 
\begin{equation}
\begin{aligned}
\psi_{11} &= \frac{k}{(k^2+1)^2} \theta_{11},  &  \psi_{01} = -\frac{3k}{4 \sigma} \psi_{11} \psi_{12}, \\
\psi_{12} &= -\frac{k}{(k^2+4)^2} \theta_{12},  &  \psi_{03} = \frac{k}{36 \sigma} \psi_{11} \psi_{12}.
\end{aligned}
\end{equation}
Under a suitable change of variables, \eqref{eq:infPr1} and \eqref{eq:infPr2} are each equivalent to the large--$\sigma$ limit of the Lorenz equations studied previously in~\cite{Stewart1989}: 
\begin{equation} \label{eq:infLorenz}
\begin{aligned}
\dot{y} &= \left(\rho - 1 - z \right) y, \\
\dot{z} &= -\beta z + y^2,
\end{aligned} 
\end{equation} 
corresponding to the restriction of Lorenz equations to the plane $x = y$. To obtain~\eqref{eq:infLorenz} from~\eqref{eq:infPr1} we change variables according to
\begin{equation}
\begin{aligned}
\theta_{11} = \sqrt{2} \R_{L_1} \, y, \qquad  \theta_{02}& = \R_{L_1} \, z, \qquad (k^2+1) t \mapsto t, \\
\beta = 4/(k^2+1)&, \qquad \rho = \R/\R_{L_1}.
\end{aligned}
\end{equation}
and a similar scaling may be applied to $\theta_{12}$ and $\theta_{04}$ to obtain~\eqref{eq:infLorenz} from~\eqref{eq:infPr2}. Therefore the dynamics of~\eqref{eq:infPr1} and~\eqref{eq:infPr2} can each be understood by studying the ODE~\eqref{eq:infLorenz}. The nonzero equilibria of \eqref{eq:infLorenz} are $(y,z) = (\pm \sqrt{\beta(\rho - 1)}, \rho - 1)$. These equilibria are globally stable within their respective half-plane ($y> 0$ or $ y < 0$) for all $\rho > 1$. In contrast to the Lorenz equations at finite $\sigma$, trajectories of \eqref{eq:infLorenz} cannot become chaotic~\cite{Stewart1989}.

When $\R > \R_{L_2}$,~\eqref{eq:infPr1} and~\eqref{eq:infPr2} each have three equilibria: a pair of Lorenz-like equilibria corresponding to the $L_1$ or $L_2$ states, and the zero equilibrium. Any combination of these provides an equilibrium for the full HK8 system, and therefore there are nine in total. The four such equilibria where both Lorenz-like systems are nonzero are stable when they exist, and correspond to the large-$\sigma$ limit of the $TC$ states. The maximum $N$ over all equilibria in the large-$\sigma$ limit is $N_{TC} - 1 = (N_{L_1} - 1) + (N_{L_2} - 1)$. It can be shown using SOS optimization with degree two auxiliary functions that for sufficiently large $\R$, the $TC$ equilibria saturate the upper bound $U^*_d$, and thus:
\begin{equation} \label{eq:infPrBound}
\max_{\mathbf{x}(t)} N = \begin{cases}
1, & 0 < \R \leq R_{L_1}, \\
3 - 2 \R_{L_1}/\R, & 2 \R_{L_1} < \R \leq \R_{L_2}, \\
5 - 2 \R_{L_1}/\R - 2 \R_{L_2}/\R, \hspace{3em} & \R > \R_{L_2}.
\end{cases}
\end{equation}
These upper bounds arise as the $\sigma \to \infty$ limit of the bounds constructed at finite $\sigma$ in~\S\ref{sec:Quad}. 
In the infinite-Prandtl number limit of the HK8 system, equilibria saturate the upper bound on $N$ for all $\R$ and $k$. This provides a contrast to the bounds at finite Prandtl number determined in~\S\ref{sec:Numerical SDP}, where time-dependent states were observed to maximize $N$ for $\R$ sufficiently large.

\subsection{TC equilibria in the infinite-$\R$ limit}

Expressions for the $TC$ equilibria can be obtained using symbolic manipulation, yet their exact formulae are too complicated to analyze directly.
In order to better understand the behavior of the $TC$ equilibria, we compute asymptotic formulae for these equilibria in the large-$\R$ limit. Solutions to the truncated model in this limit have almost no correspondence to solutions of the full PDE; the purpose of this analysis is purely to gain a better understanding of the HK8 model.

The qualitative behavior of the $TC$ states at large $\R$ may be categorized by dividing the  $k^2$--$\sigma$ plane into three distinct parameter regimes, much like the analysis performed in~\S\ref{sec:steady}. Let $S_1$ and $S_2$ be the regions where the $TC$ branch connects to only the $L_1$ or $L_2$ branch, respectively, defined by
\begin{align} \label{eq:TC_Asy_reg}
S_1 & := \{(k^2, \sigma) : (10\sigma - 3\sigma^2)(k^2+4)^2 + 2(k^2+1)(5k^2 + 11) \leq 0\}, \\
S_2 & := \{(k^2, \sigma) : (10\sigma + 3\sigma^2)(k^2+1)^2 + 2(k^2+4)(5k^2 - 4)  \leq 0\}.
\end{align}
Below, we prove that the large-$\R$ limit of $N$ is
\begin{equation}
\begin{aligned}
N_0^1 &= \frac{20(k^2+1)(5k^2+11) + 2(65k^4 + 313k^2 + 698) \sigma + 45 (k^2+4)^2 \sigma^3}{20 (k^2+1) (5k^2+11) + 2(65k^4 + 403k^2 + 788) \sigma + 9 (k^2+4)^2 \sigma^3}, & (k^2, \sigma) \in S_1, \\[6pt]
N_0^2 &= \frac{20(k^2+4)(5k^2-4) + 2(35k^4 - 83 k^2 - 442) \sigma + 45 (k^2+1)^2 \sigma^3}{20 (k^2+4) (5k^2 -4) + 2(35k^4 + 7k^2 - 82) \sigma + 9 (k^2+1)^2 \sigma^3}, &(k^2, \sigma) \in S_2.
\end{aligned}
\end{equation}
The regions $S_1$ and $S_2$ correspond with regions V and I--III, respectively, in Figure~\ref{fig:bif}. 
In the part of parameter space that separates $S_1$ and $S_2$ (called regions III--IV in Figure~\ref{fig:bif}), the $TC$ branch connects to both the $L_1$ and $L_2$ branches via pitchfork bifurcations, and as a result $TC$ equilibria only exist for a finite range of $\R$. Hence, the large-$\R$ limit of the $TC$ states need only be considered in $S_1$ and $S_2$. 

We first derive algebraic conditions on the $TC$ equilibria, using the fact that all eight variables are nonzero at the $TC$ states. Define the variables $X = \psi_{11}^2$ and $Y = \psi_{12}^2$. Then, assuming $X$ and $Y$ are nonzero, the algebraic system whose solutions are equilibria of the HK8 model reduces to
\begin{equation} \label{eq:TC_red}
\begin{aligned} 
0 &= k \left( \R_{L_1} - \mathcal{R} \right) + \frac{k}{8} (k^2+1)^2 X  + \left( (k^2+1) \alpha + \frac{5 k}{12 \sigma} (k^2+4)^2 \right) Y  + \left(\frac{k^2}{8} \alpha - \frac{5k^2}{12 \sigma} \beta \right) XY, \\
0 &= k\left( \R_{L_2} - \mathcal{R} \right) - \left((k^2+4) \beta + \frac{5k}{12\sigma} (k^2+1)^2 \right) X + \frac{k}{8} (k^2+4)^2 Y - \left( \frac{k^2}{8} \beta + \frac{5k^2}{12 \sigma} \alpha \right) XY,
\end{aligned}
\end{equation}
where 
\begin{align}
\alpha &= \frac{k(5k^2+11)}{12 \sigma^2}, &
\beta &= \frac{k(5k^2-4)}{12 \sigma^2}.
\end{align}
After further simplification, the variable $X$ in~\eqref{eq:TC_red} takes the form
\begin{equation} \label{eq:TC_X_eqn}
X = \frac{k \mathcal{R} + C_0 + C_1 Y}{D_0 + D_1 Y}, 
\end{equation}
where $Y$ solves the quadratic equation
\begin{equation} \label{eq:TC_Y_eqn}
Y^2 + (A_0 \R + A_1) Y + (B_0 \R + B_1) = 0.
\end{equation}
The constants $A_i, B_i, C_i, D_i$ are independent of $\R$ and may be determined from~\eqref{eq:TC_red}. To determine asymptotic expansions for $X$ and $Y$ as $\R \to \infty$, we let $\eps = 1/\R$ and multiply~\eqref{eq:TC_Y_eqn} by $\eps$, resulting in the singular perturbation problem
\begin{equation} \label{eq:TC_Y_eqn_eps}
\eps Y^2 + (A_0 + A_1 \eps) Y + (B_0 + B_1 \eps) = 0.
\end{equation}

\subsubsection{Outer approximation} \label{sec:Outer}

The equilibria corresponding to the infinite-$\R$ limit of the $TC$ equilibria in $S_2$ can be found by substituting the expansion $\displaystyle Y \sim \sum_n \eps^n Y_n$ into~\eqref{eq:TC_Y_eqn_eps}. This yields a hierarchy of equations for $Y_n$, with the leading term $Y_0$ given by
\begin{align} \label{eq:Y_exp_reg}
Y_0 &= -\frac{B_0}{A_0}.
\end{align}
Substituting the series for $Y$ into~\eqref{eq:TC_X_eqn} yields a geometric series with leading order
\begin{align} \label{eq:X_exp_reg}
x_0 &= \frac{k}{D_0 + D_1 Y_0}.
\end{align}
Because all eight variables must be real and nonzero, the above expansions provide limiting behavior for the $TC$ equilibria as long as $x_0, Y_0 >0$, corresponding exactly to the set $S_2$. Within this region,~\eqref{eq:Y_exp_reg} and~\eqref{eq:X_exp_reg} determine the limiting behavior for all eight variables on the $TC$ branch. If the limiting behavior of $N_{TC}$ is desired, it is more useful to express $N$ directly in terms of $X$ and $Y$, yielding an asymptotic series for $N$. After simplification, the volume-averaged expression~\eqref{eq:Nu2} for $N$ becomes
\begin{equation}
N = 1 + \frac{k}{4 \R} \left( \frac{(k^2+1)^2}{k} X + (\alpha - \beta) X Y + \frac{(k^2+4)^2}{k} Y \right).
\end{equation}
The expressions~\eqref{eq:Y_exp_reg} and~\eqref{eq:X_exp_reg} imply that the leading order behavior of $N$ in $S_2$ is given by 
\begin{equation}
\begin{aligned}
N_{0}^2 &= \frac{k}{4} \left(\frac{(k^2+1)^2}{k} x_0 + (\alpha - \beta)(x_0 Y_0)\right) \\
&= \frac{20(k^2+4)(5k^2-4) + 2(35k^4 - 83 k^2 - 442) \sigma + 45 (k^2+1)^2 \sigma^3}{20 (k^2+4) (5k^2 -4) + 2(35k^4 + 7k^2 - 82) \sigma + 9 (k^2+1)^2 \sigma^3}.
\end{aligned}
\end{equation}
Further terms in the expansion may be computed using the asymptotic series computed above. It can be shown that this expansion is valid in the set $S_2$, where the $TC$ branch connects only to $L_2$, as elsewhere the leading-order terms of $X$ or $Y$ will be negative. Within the interior of $S_2$, $N_0^2$ is strictly increasing in both $k^2$ and $\sigma$, reaching its maximum value of 3 along the entirety of the interior boundary of $S_2$, and approaching its minimum value of 1 as $\sigma$ vanishes. The leading-order term has a jump discontinuity at the point $(k^2,\sigma) = (4/5,0)$.

Convergence of the asymptotic series may be demonstrated by comparing its first few terms against computed values of $N$ at chosen parameter values. As expected, the leading-order expansion converges at a rate of $O(\eps^2)$ as $\eps \to 0$, while including more terms improves the order of accuracy. 

\subsubsection{Inner approximation} \label{sec:Inner}

The large-$\R$ limit of the $TC$ equilibria in $S_1$ can be found after rescaling~\eqref{eq:TC_Y_eqn_eps} by $y = \eps Y$, to obtain
\begin{equation}
y^2 + (A_0 + A_1 \eps) y + (B_0 \eps + B_1 \eps^2) = 0.
\end{equation}
This scaling emerges when seeking a dominant balance between the first two terms of~\eqref{eq:TC_Y_eqn_eps}. We then proceed as in~\ref{sec:Outer} to find that the leading-order expansion for $N$ is given by

\begin{equation}
\begin{aligned}
N_0^1 &= 1 + \frac{k}{4} \left((\alpha - \beta) X_0 y_0 + \frac{(k^2+4)^2}{k} y_0 \right) \\
&= \frac{20(k^2+1)(5k^2+11) + 2(65k^4 + 313k^2 + 698) \sigma + 45 (k^2+4)^2 \sigma^3}{20 (k^2+1) (5k^2+11) + 2(65k^4 + 403k^2 + 788) \sigma + 9 (k^2+4)^2 \sigma^3}.
\end{aligned}
\end{equation}
where
\begin{align}
y_0 &= -A_0, &
X_0 &= \frac{k + C_1 y_0}{D_1 y_0}.
\end{align}

This solution corresponds to the asymptotic state of the $TC$ branch in $S_1$ that is of particular interest since the $TC$ equilibria maximize $N$ among the steady states of the HK8 model at sufficiently large $\R$ in this parameter regime. The leading-order term $N_0^1$ is nearly constant in $k^2$ and is strictly increasing in $\sigma$, rapidly approaching 5 as $\sigma \to \infty$. As $(k^2, \sigma)$ approaches the interior boundary of $S_1$, the $TC$ equilibria approximate $L_1$ states and therefore $N_0^1 \to 3$ in this limit.

\section{Special case of the quadratic SOS method: results of Souza \& Doering} \label{sec:SD}
The analysis of Souza \& Doering in \cite{Souza2015} amounts to a special case of the quadratic SOS approach implemented in~\S\ref{sec:Quad}, with the coefficients of the general quadratic auxiliary function \eqref{eq:Vdef}--\eqref{eq:quadCons} restricted more than necessary. The bounds they prove are identical to those proven in \S\ref{sec:Quad} when $\R \leq \sqrt{\R_{L_1} \R_{L_2}}$, bur for larger $\R$ they prove instead that \cite{Souza2015}
\begin{equation} \label{eq:SDBound}
N \leq N_{L_1} + \frac{1}{\R}\left[ \R_{L_1} - \R_{L_2} +  \sqrt{(\R_{L_2} - \R_{L_1})^2 + 4(\R - \sqrt{\R_{L_1} \R_{L_2}})^2} \right].
\end{equation}
On this interval, more general quadratic $V$ give sharper bounds, as shown by the results of~\S\ref{sec:Numerical SDP} and~\S\ref{sec:Quad}.

To show how the analysis of~\cite{Souza2015} fits into our present framework, let us derive the large-$\R$ bound of~\eqref{eq:SDBound} in the language of \S\ref{sec:Quad}. Let $z_1$ and $z_2$ be constants to be chosen below---the ``background variables'' in the language of~\cite{Souza2015}---and define the constant
\begin{equation} \label{eq:alpha}
\alpha = \frac{z_1 + z_2}{z_1^2 + z_2^2}.
\end{equation}
Our approach in \S\ref{sec:Quad} reduces to the special case of~\cite{Souza2015} if the coefficients $c_i$ of $V$ in~\eqref{eq:Vdef} are restricted such that
\begin{equation} \label{eq:SDcoeffs}
\begin{aligned}
c_1 &= \frac{1 - 2 \alpha z_1}{2 \R}, \hspace{4em} & c_2 & = \frac{1 - 2\alpha z_2}{4 \R}, \hspace{2em} & c_3 &= \frac{(k^2+1)(\alpha - 1)}{8 \sigma \R}, \\
c_5  &= \frac{(k^2+4)(\alpha - 1)}{8 \sigma \R},  & c_8 &= \frac{\alpha}{8 \R^2}, & c_9 &= 0.
\end{aligned}
\end{equation}
With the coefficients constrained by~\eqref{eq:quadCons} and~\eqref{eq:SDcoeffs}, there are only two free parameters remaining in the expression for $V$, and therefore the auxiliary function is determined by specifying $z_1$ and $z_2$. Under these restrictions on $V$, the minimal upper bound such that the polynomial $S$ is sum-of-squares is 
\begin{equation} \label{eq:SD_N}
N \leq 1 + 2(z_1 + z_2),
\end{equation}
provided $z_1$ and $z_2$ can be chosen so that $\alpha \ge 0$ and
\begin{align}\label{eq:SD_const}
\frac{\R_{L_1}}{\R} (\alpha - 1) - \alpha (z_1 - 1)^2 &\geq 0, \\
\frac{\R_{L_2}}{\R} (\alpha - 1) - \alpha (z_2 - 1)^2 &\geq 0.
\end{align}

The bound of~\eqref{eq:SDBound} in the $\R > \sqrt{\R_{L_1} \R_{L_2}}$ case may then be constructed by taking \cite{Souza2015}
\begin{equation} \label{eq:SD_Assum_2}
\begin{aligned}
z_1 &=  \frac{ \sqrt{\R_{L_1}} \left[-\R_{L_1} + 2 \R \sqrt{\frac{\R_{L_2}}{\R_{L_1}}} - \R_{L_2} +  \sqrt{(\R_{L_2} - \R_{L_1})^2 + 4(\R - \sqrt{\R_{L_1} \R_{L_2}})^2} \right]}{2 \R \left(\sqrt{\R_{L_1}} + \sqrt{\R_{L_2}} \right)}, \\
z_2 &= \sqrt{\frac{\R_{L_2}}{\R_{L_1}}}(z_1 - 1) + 1.
\end{aligned}
\end{equation}
This bound is valid whenever ${\R > \sqrt{\R_{L_1} \R_{L_2}}}$ but is not as tight as the result obtained in~\S\ref{sec:Numerical SDP} and~\S\ref{sec:Quad} using the most general quadratic ansatz for $V$.

\section{Analytical bounds in the larger-wavenumber regime}
\label{sec:QuadBounds3}
This appendix gives partial results towards the analytical optimization of $N$ with quadratic $V$ in the regime where $\R > \R'$ and $k > k'$ (cf.\ Figure~\ref{fig:quadRegion}). In this regime the bound~\eqref{eq:quadBound2} does not hold, so the assumptions~\eqref{eq:basis2} are not valid. Our best analytical bound coincides with that of~\cite{Souza2015}:
\begin{equation} \label{eq:quadBound3}
N \leq N_{L_1} + \frac{1}{\R} \left[\R_{L_1} - \R_{L_2} + \sqrt{(\R_{L_2} - \R_{L_1})^2 + 4(\R - \sqrt{\R_{L_1} \R_{L_2}})^2} \right].
\end{equation}
The exact auxiliary function required to prove this bound is given in~\ref{sec:SD}. However, this bound is not optimal among quadratic $V$; numerical solution to \eqref{eq:Opt} with quadratic $V$ gives sharper bounds at many parameter values. 

In order for the construction of optimal $V$ to become analytically tractable, we want to further restrict $V$ in a way that will be justified \emph{a posteriori} by the sharpness of the resulting bounds, much as was done for the smaller-$k$ regime in~\S\ref{sec:QuadBounds2}. If the bounds are saturated by the $TC$ equilibria, then $S$ must vanish there. Since $\psi_{01} = -27 \psi_{03}$ on the $TC$ equilibria, 
\begin{equation} \label{eq:basis3a}
\mathbf{b}_2^T \Q_2 \mathbf{b}_2 = q_2 \left(\psi_{01} + 27 \psi_{03} \right)^2.
\end{equation}
We further observe in SOS computations with quadratic $V$ that the determinants of $\Q_1$ and $\Q_3$ are zero up to the tolerance of the solver. This implies that for some $q_1, q_2, A_1$, and $A_2$,
\begin{align} \label{eq:basis3b}
\mathbf{b}_1^\mathsf{T} \Q_1 \mathbf{b}_1 &= q_1\left( \psi_{11} - A_1 \theta_{11} \right)^2, &
\mathbf{b}_3^\mathsf{T} \Q_3 \mathbf{b}_3 &=  q_3\left(\psi_{12} - A_2 \theta_{12} \right)^2.
\end{align}
These restrictions impose the coefficient relationships
\begin{equation} \label{eq:coeffs3}
\begin{aligned}
c_3 &= \frac{(5k^2 - 4)(k^2+1)}{(5k^2+11)(k^2+4)} c_5, &
c_9 &= \frac{108}{(k^2+4)(5k^2+11)} c_5, \\
c_1 &= 8 \frac{k^2+1}{k} \sqrt{\sigma c_3 c_8} - 4 \R c_8 - \frac{4 \sigma}{k^2+1} c_3, &
c_2 &= 4 \frac{k^2+4}{k} \sqrt{\sigma c_5 c_8} - 2 \R c_8 - \frac{2 \sigma}{k^2+4} c_5.
\end{aligned}
\end{equation}
The semidefinite constraints will be satisfied if $c_5$ and $c_8$ are each nonnegative, so it remains to determine $c_5$ and $c_8$ that minimize $U$. By the same argument used in~\S\ref{sec:QuadBounds2}, the optimal $\Q_4$ must have a determinant of zero,  in which case the SOS optimization is equivalent to
\begin{equation} \label{eq:largeKOpt1}
    \min_{c_5, c_8 \geq 0} U,
\end{equation}
where
\begin{equation}
\label{eq:largeKOpt2}
\begin{aligned}
    U - 1 = \frac{1}{16 c_8} &\Bigg[ \left(8 \R c_8 + \frac{1}{\R} + \frac{8\sigma \mu}{k^2+1} c_5 - 16\frac{k^2+1}{k} (\sigma\mu  c_5 c_8)^{1/2} \right)^2 \\
    & \quad + \left(8 \R c_8 + \frac{1}{\R} + \frac{8 \sigma}{k^2+4} c_5 - 16 \frac{k^2+4}{k} (\sigma c_5 c_8)^{1/2} \right)^2 \Bigg],
\end{aligned}
\end{equation}
with 
\begin{equation} \label{eq:largeKOpt3}
    \mu = \frac{(5k^2-4)(k^2+1)}{(5k^2+11)(k^2+4)}.
\end{equation}
Numerical solutions of~\eqref{eq:largeKOpt1}--\eqref{eq:largeKOpt3}, obtained using the Mathematica function \texttt{NMinimize}, agree with numerical solutions to the full SOS optimization problem for various parameter values in this regime, suggesting the assumptions \eqref{eq:basis3a}--\eqref{eq:basis3b} leading to this simpler minimization problem are not overly restrictive. However, we have not been able to derive an analytical solution to~\eqref{eq:largeKOpt1}--\eqref{eq:largeKOpt3} that is simple enough to be useful.

\bibliography{SDP}
\bibliographystyle{abbrv}

\end{document}